\documentclass[a4paper,11pt]{article}

\usepackage{jheppub} 
\usepackage{amsmath}
\usepackage[T1]{fontenc} 
\usepackage{natbib}
\usepackage{setspace}

\def\be{\begin{equation}}
\def\ee{\end{equation}}

\def\bea{\begin{eqnarray}}
\def\eea{\end{eqnarray}}

\def\vec[#1]{\boldsymbol{#1}}
\def\vecs[#1,#2]{\boldsymbol{{#1}_{#2}}}

\def\mes[#1]{d^{3}{#1}}
\newcommand{\vect}[1]{{\boldsymbol{#1}}}
\def\del{\partial}

\def\vk{\vect{k_1}}
\def\vkk{\vect{k_2}}
\def\vkkk{\vect{k_3}}
\def\vkfour{\vect{k_4}}
\newcommand{\half}{\frac{1}{2}}

\title{Ward Identities for Scale and Special Conformal Transformations in Inflation}

\author[1]{Nilay Kundu,}
\author[2]{Ashish Shukla,}
\author[2]{and Sandip P. Trivedi}

\vspace{5mm}

\affiliation[1] {\it Harish-Chandra Research Institute, Chhatnag Road, Jhunsi,
Allahabad, 211019, India}
\affiliation[2]
{\it Department of Theoretical Physics,
 Tata Institute of Fundamental Research,\\  Colaba, Mumbai, 400005, India \\}

\vspace{1cm}

\emailAdd{nilay.tifr@gmail.com}
\emailAdd{ashukla.phy@gmail.com}
\emailAdd{trivedi.sp@gmail.com}

\abstract{We derive the general Ward identities for scale and special conformal transformations in theories of single field inflation. Our analysis is model independent and based on symmetry considerations alone.
The identities we obtain are valid to all orders in the slow roll expansion. 
For special conformal transformations, the Ward identities include a term which is non-linear in the fields that arises due to a compensating spatial reparametrization. Some observational consequences are also discussed.}


\begin{document}

\begin{flushright} 
\small{TIFR/TH/15-21 \\ HRI/ST/1509} 
\end{flushright}

\maketitle
\flushbottom

\section{Introduction}
\label{intro}
Inflation is the dominant paradigm to explain the approximate isotropy and homogeneity of the universe. It also gives rise to quantum perturbations which lead to the observed anisotropy in the Cosmic Microwave Background, and  which seed the formation of large scale structure in the universe. 
During inflation, space-time is well approximated by four dimensional de Sitter space, which is a maximally symmetric FRW cosmology, with the symmetry group $SO(1,4)$. The time evolution of the inflaton and its back-reaction on the metric breaks these symmetries, but this breaking is small if the slow roll conditions are satisfied. 

We will refer to the $SO(1,4)$ symmetry of de Sitter space as the conformal symmetry group. It includes translations and rotations along the spatial directions, as well as a scale transformation, and three special conformal transformations. 
In this paper, we explore the constraints imposed by these $SO(1,4)$ symmetries on the perturbations produced during inflation. More specifically, we derive Ward identities arising due to the scale and special conformal transformations for the correlation functions of these perturbations. Our Ward identities incorporate the breaking of the $SO(1,4)$ symmetry as well, and are valid to all orders in the slow roll parameters. 

The analysis we carry out is based on symmetries alone, and is independent of specific models. As a result, the Ward identities we obtain can provide robust model independent checks of the central idea behind a large class of inflationary models, namely, that the inflationary dynamics (including the scalar sector) preserves approximate conformal invariance. These results should apply not only to slow roll models with different shapes of the inflationary potential, but also in situations where higher derivative corrections can become important, such as in string theory scenarios, with the Hubble scale during inflation being of order the string scale, which in turn is much smaller than the Planck scale. 

The $SO(1,4)$ symmetry is also the symmetry group of a $3$-dimensional Euclidean conformal field theory, which is the motivation behind our calling it the conformal group. 
However, we should mention at the outset that we do not  assume a dS/CFT type of correspondence in deriving our results. Rather the connection with a conformal field theory (with the breaking of conformal invariance also included)
is only for the purpose of organizing our discussion of the symmetries. 

This paper is organized as follows. 
Section \ref{basics} contains the basic setup.
The central ideas and key results behind the derivation of the Ward identities are then discussed in section \ref{identities}. For our analysis, it is useful to work with the late time wave function of the universe, when the modes of interest have exited the horizon. 
Constraints imposed by symmetries on the coefficient functions determining the wave function are discussed in section \ref{conditions}. The late time behaviour of the modes in the canonical slow roll model of inflation is discussed in subsection \ref{canmodel}, and some aspects which arise when higher derivative corrections are incorporated are discussed in subsection \ref{hdcorrect}. We end with conclusions in section \ref{conclusions}. 
The three appendices contain important supplementary material. 

The analysis we carry out is based on the seminal works \cite{Maldacena:2002vr} and \cite{Maldacena:2011nz}. It also develops ideas earlier reported in \cite{Mata:2012bx}, \cite{Ghosh:2014kba} and \cite{Kundu:2014gxa}. 
There are many other references also of relevance. The use of conformal symmetry to constrain inflationary correlation functions has also been discussed in \cite{Antoniadis:1996dj, Larsen:2002et, Larsen:2003pf, McFadden:2010vh, Antoniadis:2011ib, McFadden:2011kk, Creminelli:2011mw, Bzowski:2011ab, Kehagias:2012pd, Kehagias:2012td, Schalm:2012pi, Bzowski:2012ih, McFadden:2014nta, Kehagias:2015jha}. Approaches where the conformal symmetries are often thought of as being non-linearly realized include \cite{Weinberg:2003sw, Creminelli:2004yq, Cheung:2007sv, Weinberg:2008nf,  Creminelli:2011sq, Bartolo:2011wb, Creminelli:2012ed, Hinterbichler:2012nm, Senatore:2012wy, Assassi:2012zq, Creminelli:2012qr,  Goldberger:2013rsa, Hinterbichler:2013dpa, Creminelli:2013cga, Pimentel:2013gza, Berezhiani:2013ewa, Sreenath:2014nka, Mirbabayi:2014zpa, Joyce:2014aqa, Sreenath:2014nca}. The idea of using time and spatial reparametrizations to derive Ward identities in the context of AdS was first discussed in \cite{deBoer:1999xf}
.

\textbf{Notation}:  Before proceeding, let us clarify the notation we will  follow in this paper. A  dot above a quantity represents a time derivative, e.g. ${\dot \phi} \equiv d\phi/dt $. Spatial three vectors are written in boldface, e.g. $\vec[x], \vec[k]$, etc. Also, $k_a, k_b$, etc. represent the magnitudes of the vectors $\vecs[k,a], \vecs[k,b]$, whereas $k_i, k_j$, etc. represent the $i^{th}, j^{th}$ components of  $\vec[k]$. Unless otherwise stated, the spatial indices $i,j$, etc. will be raised and lowered using the Kronecker delta, $\delta_{ij}$.

\section{Essential Ideas}
\label{basics}
In this section, we will outline the essential ideas behind the derivation of the Ward identities. Our discussion will be general and not tied to any specific model. 
In sections \ref{canmodel} and \ref{hdcorrect}, we will discuss the concrete cases of the canonical model of slow roll inflation, and the presence of higher derivatives, respectively. 

The dynamical degrees of freedom in the theories we consider will be the metric and a single scalar field  \footnote{The discussion can be extended to  include  additional scalars. However, model independent  observational predictions are not easy to make in such models.}. We work with the ADM form of the metric,  
\be
\label{admmetric}
ds^2 = - N^2 dt^2 + h_{ij} \, (dx^i + N^i dt) (dx^j + N^j dt),
\ee
with $N$ and $N^i$ being the lapse and shift functions respectively. We choose the gauge
\be
\label{gfix}
N = 1, N^i = 0.
\ee
This gauge is called the \textit{synchronous} gauge. 

The unperturbed background FRW solution is
\be
\label{frwback}
ds^2 = - \, dt^2 + a^2(t) \, \delta_{ij} \, dx^i dx^j .
\ee
The Hubble parameter is given by 
\be
\label{hubble}
H = \frac{\dot{a}}{a} .
\ee

Including the metric perturbations, denoted by $\gamma_{ij}$,  gives
\be
\label{pert_met}
h_{ij} = a^2(t) \,[\delta_{ij} + \gamma_{ij}].
\ee
Similarly, expanding the inflaton about the background value ${\bar \phi}(t)$ gives 
\be
\label{expinf}
\phi = {\bar \phi}(t) + \delta \phi.
\ee

The gauge choice, eq.\eqref{gfix}, does not fix all the coordinate reparametrization invariance. There are two kinds of residual gauge transformations which can be carried out.  These are spatial reparametrizations,
\be
\label{sprep}
x^i \rightarrow x^i + v^i(\vec[x]),
\ee
under which
\be
\label{chh}
h_{ij} \rightarrow h_{ij} + \nabla_i v_j + \nabla_j v_i,
\ee
or equivalently
\be
\label{chg}
\gamma_{ij} \rightarrow \gamma_{ij} + \frac{1}{a^2(t)} \, \big(\nabla_i v_j + \nabla_j v_i\big).
\ee

We can also perform time reparametrizations
\be
\label{timere}
t \rightarrow t + \epsilon(\vec[x]),
\ee
along with accompanying spatial reparametrizations of the form
\be
\label{xrep}
x^i \rightarrow x^i + w^i(t, \vec[x])
\ee
with
\be
\label{vix}
w^i(t, \vec[x]) = \del_i \epsilon(\vec[x]) \int^t dt' \, \frac{1}{a^2(t')} ,
\ee
under which
\be
\label{change1}
\delta\phi \rightarrow \delta\phi + \dot{\bar{\phi}}(t) \, \epsilon(\vec[x]) , 
\ee
\be
\gamma_{ij} \rightarrow \gamma_{ij} + 2\, \delta_{ij} \bigg(\frac{\dot{a}}{a}\bigg) \epsilon(\vec[x]) + \big(\del_i w_j + \del_j w_i\big). \label{change2}
\ee

Using the  homogeneity of the background FRW solution, we can expand the  perturbations in a basis of modes carrying  fixed comoving momenta. Let $\xi$ be a generic perturbation. Then
\be
\label{cm}
\xi(t,\vec[x]) = \int \frac{d^3 k}{(2\pi)^3} \, e^{i \vec[k]\cdot\vec[x]} \, \xi(t,\vec[k]),
\ee
where the comoving momentum is $\vec[k]$. 
We will be interested in the behaviour of the perturbations at late times, when the modes of interest have left the horizon,
\be
\label{lh}
k^2/a^2 \ll H^2.
\ee

Using the time reparametrization symmetry at late times, we can set 
\be
\label{delphigauge}
\delta \phi=0.
\ee
In this gauge, the perturbations freeze out once they exit the horizon, i.e., they become time independent, since their subsequent evolution becomes dominated by a frictional term proportional to the Hubble parameter.  
The remaining gauge invariance now corresponds to spatial reparametrizations, eq.\eqref{sprep}.
The choice of gauge eq.(\ref{delphigauge}), and the freeze out of modes will be discussed in greater detail for the canonical slow roll model in section \ref{canmodel}, and in the presence of higher derivative terms in section \ref{hdcorrect}.

In the gauge eq.(\ref{delphigauge}), all the remaining perturbations arise from the metric. We can decompose them as
\be
\label{decg}
\delta_{ij}+ \gamma_{ij} = e^{2\zeta} [\delta_{ij} + {\widehat \gamma}_{ij}] ,
\ee
where ${\widehat \gamma}_{ij}$ is the traceless component. 
$\zeta$ determines the perturbations in the trace of the metric. To linear order in perturbations, we see from eq.(\ref{decg}) that 
\be
\label{linorz}
\gamma_{ij} = 2\zeta \, \delta_{ij} + {\widehat \gamma}_{ij}.
\ee

Going beyond the linear order, we will find that  the definition given in eq.\eqref{decg} leads to a simplification in our discussion of symmetries.
To be more specific, it will turn out that the coefficient functions for the trace of the stress tensor will transform in a canonical way with this choice of variables.

It will be useful to carry out our symmetry based analysis in terms of the wave function of the universe.  This wave function is actually a \textit{functional} of the perturbations. 
Expanding at late times, when the perturbations become time independent, we get 
\begin{small}
\begin{equation}
\begin{split}
\Psi[\gamma_{ij}] = \text{exp} \bigg[&- \half \int \mes[x] \, \mes[y] \, \zeta(\vec[x]) \zeta(\vec[y]) \, \langle T(\vec[x]) T(\vec[y]) \rangle \\
&- \int \mes[x] \, \mes[y] \, \zeta(\vec[x]) \, \widehat\gamma_{ij}(\vec[y]) \, \langle T(\vec[x]) \widehat{T}^{ij}(\vec[y]) \rangle \\
&- \half \int \mes[x]  \, \mes[y] \, \widehat\gamma_{ij}(\vec[x]) \widehat\gamma_{kl}(\vec[y]) \, \langle \widehat{T}^{ij}(\vec[x]) \widehat{T}^{kl}(\vec[y]) \rangle \\
&- \frac{1}{3!} \int \mes[x] \, \mes[y] \, \mes[z] \, \zeta(\vec[x]) \zeta(\vec[y]) \zeta(\vec[z]) \, \langle T(\vec[x]) T(\vec[y]) T(\vec[z])\rangle \\
&- \frac{1}{2} \int \mes[x] \, \mes[y] \, \mes[z] \, \zeta(\vec[x]) \zeta(\vec[y]) \widehat\gamma_{ij}(\vec[z]) \, \langle T(\vec[x]) T(\vec[y]) \widehat{T}^{ij}(\vec[z])\rangle \\
&- \frac{1}{2} \int \mes[x] \, \mes[y] \, \mes[z] \, \zeta(\vec[x]) \widehat\gamma_{ij}(\vec[y]) \widehat\gamma_{kl}(\vec[z])\, \langle T(\vec[x]) \widehat{T}^{ij}(\vec[y]) \widehat{T}^{kl}(\vec[z]) \rangle \\
&- \frac{1}{3!} \int \mes[x] \, \mes[y] \, \mes[z] \, \widehat\gamma_{ij}(\vec[x]) \widehat\gamma_{kl}(\vec[y]) \widehat\gamma_{mn}(\vec[z]) \, \langle \widehat{T}^{ij}(\vec[x]) \widehat{T}^{kl}(\vec[y]) \widehat{T}^{mn}(\vec[z])\rangle \\ &+ \cdots \, \cdots \\ &- \frac{1}{m! \, n!} \int \mes[x_1] \cdots \mes[x_{m+n}] \, \zeta(\vecs[x,1]) \cdots \zeta(\vecs[x,m])\,\widehat\gamma_{i_1 j_1}(\vecs[x,m+1]) \cdots \widehat\gamma_{i_n j_n}(\vecs[x,m+n]) \times \\ & \hspace{30mm} \big\langle T(\vecs[x,1]) \cdots T(\vecs[x,m]) \widehat{T}^{i_1 j_1}(\vecs[x,m+1]) \cdots \widehat{T}^{i_n j_n}(\vecs[x,m+n]) \big\rangle + \cdots \bigg] .
\label{wfunction}
\end{split}
\end{equation}
\end{small}
The quadratic terms in $\zeta$ and $\widehat\gamma_{ij}$ correspond to a Gaussian wave function; higher order terms give rise to non-Gaussianity. 

Invariance with respect to the residual gauge invariance, namely with respect to the spatial reparametrization eq.\eqref{sprep}, imposes constraints on the coefficient functions $\langle T(\vec[x]) T(\vec[y]) \rangle$, $\langle {\hat T}^{ij}(\vec[x]) {\hat T}^{kl}(\vec[y]) \rangle$ etc,  which appear in this expansion. 
In  fact,  these coefficient functions have been written in a suggestive manner because the constraints take the form of Ward identities which are satisfied by correlation functions of the stress-energy tensor in a conformal field theory. 
This will be discussed further in section \ref{conditions}.
Note that in eq.\eqref{wfunction} we have also included a mixed term between $\zeta$ and ${\hat \gamma}_{ij}$ for generality, although such a term will vanish on further gauge fixing the spatial reparametrization invariance suitably, as we will see later. 
Let us also mention that 
as per our conventions, eq\eqref{linorz}, $T$ is related to the trace of the stress tensor $T_{ij}$ by 
\be
\label{relt}
T = 2 T_{ii} \equiv 2 \mathcal{T} ,
\ee
so that the coefficient function for a general metric perturbation, $\gamma_{ij}$, is $T^{ij}$. Also, $\widehat{T}_{ij}$ is the traceless part of the stress-energy tensor $T_{ij}$.

The invariance with respect to spatial reparametrizations eq.\eqref{sprep} arises as follows. The wave function as a functional of the late time value for a generic perturbation $\xi$ can be written as a path integral
\be
\label{actdp}
\Psi[\xi] = \int_{initial}^{\,\xi} \, [\mathcal{D}\xi] \, e^{i S} ,
\ee
where the initial conditions will be taken to be the Bunch-Davies vacuum. 
The action $S$ has a pre-factor $1/G \sim M_{Pl}^2$. By suitably rescaling fields in terms of the Hubble parameter $H$, we see that 
\be
\label{ress}
S = {M_{Pl}^2 \over H^2} \, {\tilde S},
\ee
where ${\tilde S}$ contains the rescaled fields which have been made dimensionless by the rescaling. 

Since no gravity waves have been detected so far, we know that \footnote{We take $M_{Pl} = {1\over \sqrt{8 \pi G}} \approx 10^{18} \text{GeV}$.}
\be
\label{bh}
{H^2\over M_{Pl}^2}\le 10^{-8} .
\ee
Thus the path integral on the RHS of eq.\eqref{actdp} can be evaluated in the semi-classical limit, by solving the equations of motion subject to the boundary conditions at late and early times. 
In particular, in the gauge eq.\eqref{gfix}, the $N, N^i$ equations must also be imposed. These equations give rise to the invariance of the wave function under spatial reparametrizations, eq.\eqref{sprep}, after fixing the gauge, eq.\eqref{delphigauge}, at late times. 

\section{Ward Identities}
\label{identities}
We are now ready to discuss the derivation of the Ward identities. We will be interested in the Ward identities which arise due to scale and special conformal transformations. 
It is useful to first consider the case of de Sitter space, with the background metric
\be
\label{dsb}
ds^2 = - \, dt^2 + e^{2Ht} \delta_{ij} dx^i dx^j .
\ee
This metric is well known to have an $SO(1,4)$ symmetry with ten generators. Besides the three spatial translations, and three rotations along the spatial directions, this symmetry group includes scale transformations,
\be
\label{stran}
x^i \rightarrow \lambda x^i, \, t \rightarrow t - \frac{1}{H} \, \text{log}(\lambda),
\ee
and three special conformal transformations,
\begin{equation}
\begin{split}
\label{sctsymm}
x^i &\rightarrow x^i - 2(b_j x^j) x^i + b^i \bigg(\sum_{j}\, (x^j)^2 - \frac{1}{H^2} e^{-2Ht}\bigg),\\
t &\rightarrow t + \frac{2b_jx^j}{H}.
\end{split}
\end{equation}

The scale and special conformal  symmetries give rise to Ward identities on the  correlation functions of the perturbations. In de Sitter space these identities are met exactly; in inflationary backgrounds, there are corrections that arise due to the evolving inflaton which breaks these symmetries. We will derive the resulting identities for the correlation functions to all orders in the slow roll parameters
\begin{equation}
\label{epsdef}
\epsilon_1 = - \, \frac{\dot{H}}{H^2},
\end{equation}
\be
\label{deltadef}
\delta = \frac{\ddot{H}}{2 H \dot{H}},
\ee
and
\be
\label{eps1}
\epsilon = \half \frac{\dot{\bar{\phi}}^2}{H^2}.
\ee
The identities for the scale transformations are the analogues of the Callan-Symanzik equations in field theory, which incorporate the running of the coupling constants.
Similarly, we get identities for the special conformal transformations also incorporating the evolving inflaton. 

Before proceeding, let us note that in the canonical slow roll model, discussed in section \ref{canmodel}, $\epsilon$ and $\epsilon_1$ are related as
\be
\label{reln}
\epsilon = \epsilon_1.
\ee
But more generally, when higher derivatives are included, they will not be related in this way.
Also, for the slow roll conditions to hold, 
\be
\label{slrcond}
\epsilon_1, \delta, \epsilon \ll 1.
\ee

The expectation values for  the perturbations  are obtained from the wave function in the standard manner.
For example, for scalar perturbations $\zeta$ these are given by
\be
\label{zneval1}
\langle \zeta(\vecs[x,1]) \cdots \zeta(\vecs[x,n])\rangle = \frac{1}{\mathcal{N}} \int [\mathcal{D}\zeta] \, [\mathcal{D}\widehat{\gamma}_{ij}] \, |\Psi|^2 \, \zeta(\vecs[x,1]) \cdots \zeta(\vecs[x,n]),
\ee
where $\mathcal{N}$ denotes the overall normalization factor in the path integral,
\be
\label{normdef1}
\mathcal{N} = \int [\mathcal{D}\zeta] \, [\mathcal{D}\widehat{\gamma}_{ij}] \, |\Psi|^2.
\ee
We will be interested in calculating these expectation values at late times, when the perturbations of interest have frozen out. 

There is one important point which we must consider before we proceed. The sum over all metric perturbations $\widehat\gamma_{ij}$ on the RHS of eq.\eqref{zneval1} is ill defined because  we have not yet fixed the spatial reparametrization invariance symmetry. 
The integral on the RHS would diverge without fixing this symmetry. A conventional choice, which we will also make, is to take $\widehat\gamma_{ij}$ to be transverse,
\be
\label{transverse}
\del_i \widehat\gamma_{ij} = 0,
\ee
besides also being traceless.
With this further gauge fixing, the path integral on the RHS of eq.\eqref{zneval1} becomes finite. Note that since  $\widehat\gamma_{ij}$ freezes out at late times,
the additional gauge fixing required for eq.\eqref{transverse} can be achieved by a spatial reparametrization $x^i \rightarrow x^i + \epsilon^i(\vec[x])$ which preserves the synchronous gauge eq.\eqref{gfix}.

Also note that after the additional gauge fixing, eq.\eqref{transverse}, the resulting perturbations manifestly correspond to a scalar $\zeta$ with spin $0$, and a tensor perturbation ${\widehat \gamma}_{ij}$ which has spin $2$, with respect to the rotations along the spatial directions. 
 
It is worth commenting here that the underlying reason for this further gauge fixing is that we are working with local correlation functions in a theory of quantum gravity. These correlations are well defined perturbatively about the inflationary background, but only after gauge fixing, as discussed above.

 \subsection{Ward Identities for Scale Transformations}
 \label{scaletr}
 Under a scale transformation
 \be
 \label{scalechn}
 x^i \rightarrow x^i + \lambda x^i, \, \lambda \ll 1,
 \ee
 $\zeta$ and ${\widehat \gamma_{ij}}$ transform as
 \be
 \zeta \rightarrow \zeta + \lambda + \lambda \, x^i \del_i \zeta \label{zchscal1} \,,
 \ee
 \be
 \widehat\gamma_{ij} \rightarrow \widehat\gamma_{ij} + \lambda \, x^k \del_k \widehat\gamma_{ij} \label{gchscal1}
 \ee
 (see appendix \ref{perttrans}).
 Note that the transversality condition, eq.\eqref{transverse}, is preserved by this transformation. 
 
 We can now consider changing variables in the path integral on the RHS of eq.\eqref{zneval1}, with $\zeta$ and ${\widehat \gamma_{ij}}$ transforming as given in eq.\eqref{zchscal1} and eq.\eqref{gchscal1}, respectively. 
 The measure in the path integral on the RHS of eq.\eqref{zneval1} is invariant under spatial reparametrizations, and therefore under the change in eq.\eqref{scalechn}.

 Naively, on the basis of what has been discussed so far,  one might also conclude  that the wave function $\Psi$ is invariant under this transformation, leading to the condition
 \begin{equation}
 \label{news}
 \big\langle \delta(\zeta(\vecs[x,1])) \cdots \zeta(\vecs[x,n])\big\rangle + \cdots + \big\langle \zeta(\vecs[x,1]) \cdots \delta(\zeta(\vecs[x,n]))\big\rangle = 0,
 \end{equation}
 where from eq.\eqref{zchscal1},
 \be
 \label{defdz}
 \delta \zeta = \lambda + \lambda x^i \del_i \zeta.
 \ee
 This is incorrect.

 Under the transformation eq.\eqref{scalechn}, ${\widehat \gamma_{ij}}$ transforms homogeneously, but $\zeta$ has a homogeneous and an inhomogeneous term in its transformation. The relations between the coefficient functions we obtain, as discussed in section \ref{conditions},  will ensure that terms in the wave function which are quadratic  or higher order in the perturbations cancel amongst each other under this transformation. However, there is one term which arises from the leading term that is quadratic in $\zeta$ in eq.\eqref{wfunction}, to begin with,  which needs to be handled with care and does not cancel, as is also discussed at the end of section \ref{wscale}.
 The quadratic terms in the wave function include 
 \be
 \label{qw}
 \Psi \sim \text{exp} \bigg(- \half \int \mes[x] \, \mes[y] \, \zeta(\vec[x]) \zeta(\vec[y]) \, \langle T(\vec[x]) T(\vec[y]) \rangle \bigg).
 \ee
 After the transformation eq.\eqref{zchscal1},
 we get a piece arising from the inhomogeneous term in the transformation of $\zeta$,
 \be
 \label{zinh}
 \zeta \rightarrow \zeta + \lambda + \cdots  ,
 \ee
 which will now be linear in $\zeta$,
 \be
 \label{linp}
 \delta\Psi \sim \text{exp} \bigg(- \lambda \int \mes[x] \, \mes[y] \, \zeta(\vec[x]) \, \langle T(\vec[x]) T(\vec[y]) \rangle \bigg).
 \ee
 This term will remain uncanceled. In contrast, the homogeneous term in the transformation of $\zeta$,
 \be
 \label{zhom}
 \zeta \rightarrow \zeta + \lambda x^i \del_i \zeta + \cdots  ,
 \ee
 will give rise to a term which is quadratic in $\zeta$; this will cancel against a term coming from the piece of $\Psi$ cubic  in $\zeta$.
 
 Before proceeding, let us note that in eq.\eqref{wfunction} there is another quadratic term,
 \be
 \label{ext}
 \Psi \sim \text{exp} \bigg(- \int \mes[x] \, \mes[y] \, \zeta(\vec[x]) \widehat\gamma_{ij}(\vec[y]) \, \langle T(\vec[x]) \widehat{T}^{ij}(\vec[y]) \rangle \bigg),
 \ee
 involving both $\zeta$ and ${\widehat \gamma_{ij}}$, which could also have potentially contributed an additional piece. However, in the gauge eq.\eqref{transverse}, this term in the wave function vanishes. This follows after noting that symmetries require the momentum space coefficient function $\langle \widehat{T}^{ij}(\vecs[k,1]) T(\vecs[k,2])\rangle$ to be of the form
\be
\label{gencon}
\langle \widehat{T}^{ij}(\vecs[k,1]) T(\vecs[k,2])\rangle \sim (2\pi)^3 \delta^3(\vecs[k,1] + \vecs[k,2]) \, \bigg(\frac{1}{3} \, \delta_{ij} - \frac{k_{1i} \, k_{1j}}{k_1^2} \bigg) \beta(k_1),
\ee
where $\beta(k_1)$ is a dimension 3 function of $k_1$.
 
 Keeping this uncanceled linear term, eq.\eqref{linp}, gives us then the correct Ward identity
 \begin{equation}
 \begin{split}
 \label{newa}
 \big\langle \delta(\zeta(\vecs[x,1])) \cdots \zeta(\vecs[x,n])\big\rangle &+ \cdots + \big\langle \zeta(\vecs[x,1]) \cdots \delta(\zeta(\vecs[x,n]))\big\rangle \\ &= 2 \lambda \int d^3x \, d^3y \, \langle T(\vec[x]) T (\vec[y]) \rangle \, \langle \zeta(\vecs[x,1]) \cdots \zeta(\vecs[x,n]) \zeta(\vec[x])\rangle.
 \end{split}
 \end{equation}
 We will be interested in the expectation values for $\zeta$ with non-zero momentum. Since $\lambda$ is a constant, we can drop the piece linear in $\lambda$ on the LHS of eq.\eqref{newa}, leading to the Ward identity
 \begin{equation}
\label{wardsc1}
\bigg(\sum_{\textbf{a} = 1}^n \vecs[x,a] \cdot \frac{\del}{\del \vecs[x,a]}\bigg) \langle \zeta(\vecs[x,1]) \cdots \zeta(\vecs[x,n]) \rangle = 2 \int d^3x \, d^3y \, \langle T(\vec[x]) T (\vec[y]) \rangle \, \langle \zeta(\vecs[x,1]) \cdots \zeta(\vecs[x,n]) \zeta(\vec[x])\rangle.
\end{equation}
Expressing this in momentum space gives \footnote{A prime $'$ symbol on a correlation function denotes the suppression of the overall momentum conserving delta function. For e.g.
\begin{equation*}
 \langle \zeta(\vecs[k,1]) \zeta(\vecs[k,2]) \rangle = (2\pi)^3 \delta^3(\vecs[k,1]+\vecs[k,2]) \, \langle \zeta(\vecs[k,1]) \zeta(\vecs[k,2]) \rangle'.  
\end{equation*}
}
\begin{equation}
\begin{split}
\label{wardsc2}
\bigg( 3(n-1) + \sum_{a=1}^n k_a \, &\frac{\del}{\del k_a} \bigg) \langle \zeta(\vecs[k,1]) \cdots \zeta(\vecs[k,n]) \rangle' = \\ &- \, \frac{1}{\langle \zeta(\vecs[k,n+1]) \zeta(-\vecs[k,n+1])\rangle'} \, \langle \zeta(\vecs[k,1]) \cdots \zeta(\vecs[k,n+1]) \rangle'\bigg|_{\vecs[k,n+1] \rightarrow 0}.
\end{split}
\end{equation} 
 
 Similarly, for correlation functions of tensor perturbations $\widehat\gamma_{ij}$ we get
 \begin{equation}
\begin{split}
\label{wardsc3}
\bigg(\sum_{\textbf{a} = 1}^n \vecs[x,a] \cdot \frac{\del}{\del \vecs[x,a]}\bigg) \langle &\widehat\gamma_{i_1 j_1}(\vecs[x,1]) \cdots \widehat\gamma_{i_n j_n}(\vecs[x,n]) \rangle = \\ &2 \int d^3x \, d^3y \, \langle T(\vec[x]) T (\vec[y]) \rangle \, \langle \widehat\gamma_{i_1 j_1}(\vecs[x,1]) \cdots \widehat\gamma_{i_n j_n}(\vecs[x,n]) \zeta(\vec[x])\rangle,
\end{split}
\end{equation}
 which in momentum space takes the form
 \begin{equation}
 \begin{split}
 \label{wardsc4}
 \bigg( 3(n-1) &+ \sum_{a=1}^n k_a \, \frac{\del}{\del k_a} \bigg) \langle \widehat\gamma_{i_1 j_1}(\vecs[k,1]) \cdots  \widehat\gamma_{i_n j_n}(\vecs[k,n]) \rangle' = \\ &- \, \frac{1}{\langle \zeta(\vecs[k,n+1]) \zeta(-\vecs[k,n+1])\rangle'} \, \langle \widehat\gamma_{i_1 j_1}(\vecs[k,1]) \cdots  \widehat\gamma_{i_{n} j_{n}}(\vecs[k,n]) \zeta(\vecs[k,n+1])\rangle'\bigg|_{\vecs[k,n+1] \rightarrow 0}.
 \end{split}
 \end{equation}
 Mixed identities involving both tensor and scalar perturbations can also be similarly obtained. These are given by
 \begin{equation}
 \begin{split}
 \label{wardsc5}
 &\bigg( 3(n-1) + \sum_{a=1}^n k_a \, \frac{\del}{\del k_a} \bigg) \langle \widehat\gamma_{i_1 j_1}(\vecs[k,1]) \cdots  \widehat\gamma_{i_m j_m}(\vecs[k,m]) \zeta(\vecs[k,m+1]) \cdots \zeta(\vecs[k,n])\rangle' \\ &= \, - \, \frac{1}{\langle \zeta(\vecs[k,n+1]) \zeta(-\vecs[k,n+1])\rangle'} \,  \langle \widehat\gamma_{i_1 j_1}(\vecs[k,1]) \cdots  \widehat\gamma_{i_m j_m}(\vecs[k,m]) \zeta(\vecs[k,m+1]) \cdots \zeta(\vecs[k,n+1]) \rangle'\bigg|_{\vecs[k,n+1] \rightarrow 0}.
 \end{split}
 \end{equation}
 
 Equations \eqref{wardsc2} and \eqref{wardsc4} are examples of  Maldacena consistency conditions in the literature \cite{Maldacena:2002vr}. These are exact to all orders in the slow roll expansion. 
 
 The physical picture behind these relations is easy to state. The LHS of eq.\eqref{wardsc1} is the change of the $n$-point correlator under an overall change of scale. 
 Exactly such a transformation is generated by a  scalar perturbation $\zeta(\vecs[k,n+1])$ in the limit of very long wavelength, $k_{n+1} \rightarrow 0$, leading to the identity eq.\eqref{wardsc2}.
   
   {\it  Comments}:
   The reader will note that the scale transformation eq.\eqref{scalechn} is different from the isometry in de Sitter space, eq.\eqref{stran}. 
   In de Sitter space, metric perturbations and also perturbations for test scalars freeze out at late times and become time independent. 
   Thus, in effect, the scale transformation becomes eq.\eqref{scalechn}. In the inflationary case, once we choose the gauge where eq.\eqref{delphigauge} is met, we cannot make any time reparametrization, eq.\eqref{timere}.
   Thus the only symmetries available are spatial reparametrizations. 
   
   Similarly, the special conformal transformations, which we will consider next, eq.\eqref{sctn}, are different from the corresponding isometries in de Sitter space, eq.\eqref{sctsymm}. However, again at late times, their action on time independent fields will be the same as eq.\eqref{sctn}. 
   
   It is also worth emphasizing that our derivation of the Ward identities for scale invariance obtained here  is quite general. As mentioned above, it is valid to all orders in the slow roll expansion, and thus should hold even when the slow roll conditions are not valid. 
   The assumptions one has used are that one can go to the gauge eq.\eqref{delphigauge}, and that the remaining metric perturbations in this gauge then  freeze out  due to the cosmological expansion.
   The residual spatial reparametrizations are then enough to give rise to the Ward identities above. A similar comment will also apply to the Ward identities of special conformal invariance we derive next.

 \subsection{Ward Identities for Special Conformal Transformations}
 \label{sctr}
 We next turn to the special conformal transformations,
 \be
\label{sctn}
x^i \rightarrow x^i + \alpha^i(\vec[x]) , \, \alpha^i(\vec[x]) = - 2 (\vec[b] \cdot \vec[x]) x^i + b^i \vec[x]^2.
\ee
 Here there is an important extra subtlety. Consider the transformation of $\zeta$ and $\widehat\gamma_{ij}$ under eq.\eqref{sctn} (see appendix \ref{perttrans}),
\be
\zeta(\vec[x]) \rightarrow \zeta(\vec[x]) - 2 (\vec[b] \cdot \vec[x]) + \alpha^i \del_i \zeta (\vec[x])\,,
\label{sctzc}
\ee
\be
\widehat\gamma_{ij}(\vec[x]) \rightarrow \widehat\gamma_{ij}(\vec[x]) + \alpha^m \del_m \widehat\gamma_{ij}(\vec[x]) + 2 \, \mathcal{M}_{im}^{\vec[b]}(\vec[x]) \widehat\gamma_{jm}(\vec[x]) + 2 \, \mathcal{M}_{jm}^{\vec[b]}(\vec[x]) \widehat\gamma_{im}(\vec[x])  \label{sctgc},
\ee
where $\mathcal{M}_{ij}^{\vec[b]}(\vec[x])$ is given by
\be
\label{defmop}
\mathcal{M}_{ij}^{\vec[b]}(\vec[x]) = x_i b_j - x_j b_i .
\ee
It is easy to see that the transformation eq.\eqref{sctgc} does not preserve the transverse gauge condition eq.\eqref{transverse} we have chosen for ${\widehat \gamma_{ij}}$. 
 We must therefore carry out a compensating coordinate transformation 
 \begin{equation}
 \begin{split}
 \label{compens}
 x^i &\rightarrow x^i + v^i(\vec[x]), \\ v^i(\vec[x]) &= -\,  \frac{6 b^m \widehat\gamma_{im}(\vec[x])}{\del^2},
 \end{split}
 \end{equation}
 which then restores the transversality condition on ${\widehat \gamma_{ij}}$.  Under this compensating transformation, $\zeta$ and ${\widehat \gamma_{ij}}$ transform as 
 \be
\label{zchgcom}
\zeta(\vec[x]) \rightarrow \zeta(\vec[x]) - \frac{6 b^m \widehat\gamma_{km}(\vec[x])}{\del^2} \, \del_k\zeta(\vec[x]) - \frac{2b^m \del_i \widehat\gamma_{jm}(\vec[x])}{\del^2} \, \widehat\gamma_{ij}(\vec[x]) ,
\ee
and
\begin{equation}
\begin{split}
\label{cochgf}
\widehat\gamma_{ij}(\vec[x]) \rightarrow \widehat\gamma_{ij}(\vec[x]) &- 6 b^m \left[\partial_i \bigg( \frac{\widehat\gamma_{jm}(\vec[x])}{\del^2} \bigg) + \partial_j \bigg(\frac{\widehat\gamma_{im}(\vec[x])}{\del^2} \bigg)\right]\\ & -6b^m \left[\widehat\gamma_{ik}(\vec[x]) \, \partial_j \bigg({\widehat\gamma_{km}(\vec[x]) \over \partial^2}\bigg)  + \widehat\gamma_{jk}(\vec[x]) \, \partial_i \bigg({\widehat\gamma_{km}(\vec[x]) \over \partial^2}\bigg)   \right] \\
&-6b^m \del_k \widehat\gamma_{ij}(\vec[x]) \, \bigg( \frac{\widehat\gamma_{km}(\vec[x])}{\del^2} \bigg) \\&+ 4b^m \, \widehat\gamma_{ab}(\vec[x]) \, \partial_a \bigg({\widehat\gamma_{bm}(\vec[x]) \over \partial^2}\bigg) \big(\delta_{ij} + \widehat\gamma_{ij}(\vec[x])\big) .
\end{split}
\end{equation}
 The Ward identities then arise because of the combined transformations eq.\eqref{sctzc} and eq.\eqref{zchgcom} for the transformation of $\zeta$, and eq.\eqref{sctgc} and eq.\eqref{cochgf} for the transformation of $\widehat\gamma_{ij}$. 
 Note that the compensating transformation parameter $v^i$ itself depends on ${\widehat \gamma_{ij}}$. As a result, the compensating  transformation becomes non-linear in the perturbations. 
 
Once this subtlety requiring a compensating coordinate transformation is taken care of, the rest of the analysis follows along similar lines to that for the scale transformation case. 
The wave function $\Psi$ is invariant under spatial reparametrizations, and therefore under the combined transformations eq.\eqref{sctn} and eq.\eqref{compens}. 
More correctly, this is true for all terms in the wave function which are quadratic or higher order in the perturbations. However, the inhomogeneous term in eq.\eqref{sctzc}, $-2(\vec[b]\cdot\vec[x])$, gives rise to a term in the change of the wave function which is linear in $\zeta$.
This term does not cancel. 
As a result we get a Ward identity for scalar perturbations of the form 
\begin{equation}
\begin{split}
\label{wardt1}
4 \int d^3x \, d^3y \, &(\vec[b]\cdot\vec[x]) \langle T(\vec[x]) T (\vec[y]) \rangle \langle \zeta(\vecs[x,1]) \cdots \zeta(\vecs[x,n]) \zeta(\vec[y])\rangle \\ &+ \langle \delta^C\zeta(\vecs[x,1]) \cdots \zeta(\vecs[x,n]) \rangle + \cdots + \langle \zeta(\vecs[x,1]) \cdots \delta^C\zeta(\vecs[x,n]) \rangle = 0,
\end{split}
\end{equation}
where $\delta^C \zeta$ denotes the complete homogeneous change in $\zeta$ under eq.\eqref{sctzc} and eq.\eqref{zchgcom},
\begin{equation}
\label{zcomchg}
\delta^C\zeta(\vec[x]) = \big(-2(\vec[b]\cdot\vec[x]) x^k + b^k \vec[x]^2\big)\,\del_k \zeta(\vec[x]) - \frac{6 b^m \widehat\gamma_{km}(\vec[x])}{\del^2} \, \del_k\zeta(\vec[x]) - \frac{2b^m \del_i \widehat\gamma_{jm}(\vec[x])}{\del^2} \, \widehat\gamma_{ij}(\vec[x]).
\end{equation}
In momentum space this takes the form
\begin{equation}
\begin{split}
\label{wardt2}
\big\langle \delta(\zeta(\vecs[k,1])) \cdots \zeta(\vecs[k,n]) \big\rangle + \cdots &+ \big\langle \zeta(\vecs[k,1]) \cdots \delta(\zeta(\vecs[k,n])) \big\rangle = \\ &- 2 \bigg( \vec[b]\cdot\frac{\del}{\del \vecs[k,n+1]}\bigg) \frac{\langle \zeta(\vecs[k,1]) \cdots \zeta(\vecs[k,n+1]) \rangle}{\langle \zeta(\vecs[k,n+1]) \zeta(-\vecs[k,n+1])\rangle'} \, \bigg|_{\vecs[k,n+1] \rightarrow 0},
\end{split}
\end{equation}
where $\delta(\zeta(\vec[k]))$ is given by
\begin{equation}
\begin{split}
\label{comchz}
\delta(\zeta(\vec[k])) = \widehat{\mathcal{L}}^{\,\vec[b]}_{\vec[k]} \, \zeta(\vec[k]) &+ 6\, b^m k^i \int \frac{d^3\tilde{k}}{(2\pi)^3}\, \frac{1}{\tilde{k}^2}\, \zeta(\vec[k] - \tilde{\vec[k]}) \, \widehat\gamma_{im}(\tilde{\vec[k]}) \\ &+ 2 \, b^m k^i \int \frac{d^3\tilde{k}}{(2\pi)^3}\, \frac{1}{\tilde{k}^2} \, \widehat\gamma_{ij}(\vec[k] - \tilde{\vec[k]}) \, \widehat\gamma_{jm}(\tilde{\vec[k]}),
\end{split}
\end{equation}
and the operator $\widehat{\mathcal{L}}^{\,\vec[b]}_{\vec[k]}$ is given by
\be
\label{deflh}
\widehat{\mathcal{L}}^{\,\vec[b]}_{\vec[k]} = 2 \, \Big( \vec[k] \cdot \frac{\del}{\del \vec[k]}\Big) \Big( \vec[b] \cdot \frac{\del}{\del \vec[k]} \Big) - (\vec[b]\cdot\vec[k]) \Big( \frac{\del}{\del \vec[k]} \cdot \frac{\del}{\del \vec[k]} \Big) + 6 \bigg( \vec[b] \cdot \frac{\del}{\del \vec[k]} \bigg).
\ee

Similarly, for the tensor perturbations we get the Ward identity
\begin{equation}
\begin{split}
\label{wardt3}
\big\langle {\delta} \big(\widehat\gamma_{i_1 j_1}(\vec[k_1])\big) &\cdots \widehat\gamma_{i_n j_n}(\vec[k_n]) \big\rangle + \cdots+ \big\langle \widehat\gamma_{i_1 j_1}(\vec[k_1]) \cdots {\delta} \big(\widehat\gamma_{i_n j_n}(\vec[k_n])\big) \big\rangle \\
& = -\, 2\, \bigg(\vec[b]\cdot\frac{\del}{\del \vecs[k,n+1]}\bigg) {\big\langle\widehat\gamma_{i_1 j_1}(\vec[k_1]) \cdots \widehat\gamma_{i_n j_n}(\vec[k_n]) \zeta(\vec[k_{n+1}]) \big\rangle \over \langle\zeta(\vec[k_{n+1}]) \zeta(-\vec[k_{n+1}]) \rangle'} \Bigg|_{\vec[k_{n+1}]\rightarrow 0} ,
\end{split}
\end{equation}
where ${\delta} (\widehat\gamma_{ij})$ is the complete change in $\widehat\gamma_{ij}$ in momentum space, given by

\begin{equation}
\begin{split}
\label{comchgg}
{\delta} \big(\widehat\gamma_{ij} (\vec[k])\big) = &\widehat{\mathcal{L}}^{\, \vec[b]}_{\vec[k]} \, \widehat\gamma_{ij}(\vec[k]) + 2 \, \tilde{\mathcal{M}}_{im}^{\vec[b]} (\vec[k]) \, \widehat\gamma_{jm}(\vec[k]) + 2 \, \tilde{\mathcal{M}}_{jm}^{\vec[b]}(\vec[k]) \,\widehat\gamma_{im}(\vec[k]) \\
&+ 6 b^m \, \frac{1}{k^2} \, \big(k_i \, \widehat\gamma_{jm}(\vec[k]) + k_j \, \widehat\gamma_{im}(\vec[k])\big) \\
&+ 6b^m \int \frac{d^3 k'}{(2\pi)^3} \, \frac{\widehat\gamma_{km}(\vec[k]')}{(k')^2} \, \big(  k'_i \, \widehat\gamma_{jk}(\vec[k] - \vec[k]') + k'_j \, \widehat\gamma_{ik}(\vec[k] - \vec[k]')\big) \\
&+ 6b^m \int \frac{d^3 k'}{(2\pi)^3} \, \frac{k'_l}{|\vec[k] - \vec[k]'|^2}\, \widehat\gamma_{ij}(\vec[k]') \, \widehat\gamma_{lm}(\vec[k] - \vec[k]') \\
&- 4b^m \delta_{ij} \int \frac{d^3 k'}{(2\pi)^3} \, \frac{k_a}{|\vec[k] - \vec[k]'|^2}\, \widehat\gamma_{ab}(\vec[k]') \, \widehat\gamma_{bm}(\vec[k] - \vec[k]') \\
&- 4 b^m \int \frac{d^3 k'}{(2\pi)^3} \frac{d^3 k''}{(2\pi)^3} \, \frac{(k_a - k'_a)}{|\vec[k] - \vec[k]' - \vec[k]''|^2} \, \widehat\gamma_{ij}(\vec[k]') \, \widehat\gamma_{ab}(\vec[k]'') \, \widehat\gamma_{bm}(\vec[k] - \vec[k]' - \vec[k]''),
\end{split}
\end{equation}
where
\be
\label{defmt}
\tilde{\mathcal{M}}^{\vec[b]}_{ij}(\vec[k]) = b_j \, \frac{\del}{\del k^i} - b_i \, \frac{\del}{\del k^j} \,.
\ee

Finally, we can write the general Ward identity for the variation of a mixed correlator involving $m$ tensor perturbations $\widehat\gamma_{ij}(\vec[k])$ and $(n-m)$ scalar perturbations $\zeta(\vec[k])$ as
\begin{equation}
\begin{split}
\label{finwardn}
&\big\langle {\delta} \big(\widehat\gamma_{i_1 j_1}(\vec[k_1])\big) \cdots \widehat\gamma_{i_m j_m}(\vec[k_m]) \, \zeta(\vecs[k,m+1]) \cdots \zeta(\vecs[k,n]) \big\rangle + \cdots \\ &+ \big\langle \widehat\gamma_{i_1 j_1}(\vec[k_1]) \cdots {\delta} \big(\widehat\gamma_{i_m j_m}(\vec[k_m])\big) \, \zeta(\vecs[k,m+1]) \cdots \zeta(\vecs[k,n]) \big\rangle \\
&+ \big\langle \widehat\gamma_{i_1 j_1}(\vec[k_1]) \cdots \widehat\gamma_{i_m j_m}(\vec[k_m]) \, \delta(\zeta(\vecs[k,m+1])) \cdots \zeta(\vecs[k,n]) \big\rangle + \cdots\\
&+ \big\langle \widehat\gamma_{i_1 j_1}(\vec[k_1]) \cdots \widehat\gamma_{i_m j_m}(\vec[k_m]) \,\zeta(\vecs[k,m+1]) \cdots \delta(\zeta(\vecs[k,n])) \big\rangle\\
& = -\, 2\, \bigg(\vec[b]\cdot\frac{\del}{\del \vecs[k,n+1]}\bigg) {\big\langle\widehat\gamma_{i_1 j_1}(\vec[k_1]) \cdots \widehat\gamma_{i_m j_m}(\vec[k_m]) \zeta(\vec[k_{m+1}]) \cdots \zeta(\vec[k_{n+1}]) \big\rangle \over \langle\zeta(\vec[k_{n+1}]) \zeta(-\vec[k_{n+1}]) \rangle'} \Bigg|_{\vec[k_{n+1}]\rightarrow 0} ,
\end{split}
\end{equation}
where  $\delta(\zeta(\vec[k]))$ is given by eq.\eqref{comchz} and ${\delta}(\widehat\gamma_{ij}(\vec[k]))$ is given by eq.\eqref{comchgg}.

 These identities are again exact, like the ones for scale transformations derived in section \ref{scaletr}. They are valid to all orders in the slow roll expansion, and are one of the key results of this paper. 
 Note that due to the non-linear nature of the transformations eq.\eqref{zchgcom} and eq.\eqref{cochgf}, the resulting identities are in fact quite complicated. 
 We will see in section \ref{wardcheck}, following \cite{Ghosh:2014kba}, that the identity for the four point scalar perturbations in de Sitter space is indeed met. 
 
 It is important to note that both connected and disconnected contributions to the correlation functions may be important in the Ward identities. 
 As mentioned in eq.\eqref{ress}, after suitable rescaling the action has a factor of $M_{Pl}^2/H^2$ in front of it. Since this ratio is large, eq.\eqref{bh}, the situation in cosmology is analogous to the large $N$ limit in AdS/CFT, with $M_{Pl}^2/H^2$ playing the role of $N^2$. Disconnected components in the Ward identities can then often dominate over connected ones. More accurately, the non-linear nature of the transformation in the compensating spatial reparametrization means that different number of fields will be present in  correlation functions involved in the LHS of the Ward identities eq.\eqref{wardt2}, eq.\eqref{wardt3}. The suppression at large $N$ due to additional fields can be compensated for by including additional disconnected components. For more details and an explicit example see section \ref{wardcheck}. 
  
  Let us also note that in the Ward identity for scalar perturbations eq.\eqref{wardt2}, for the cases  $n=2, 3$,  the extra  terms in $\delta \zeta(\vec[k])$ due to the compensating spatial reparametrization can be neglected to the leading order in $H^2/M_{Pl}^2$. The extra terms are the last two terms on the RHS of eq.\eqref{comchz}. This will become clearer from the discussion in section \ref{wardcheck}.  As a result, for the cases $n=2, 3$, the Ward identity eq.\eqref{wardt2} takes the form, 
  \begin{equation}
  \label{limeq1}
 \left( \sum_{a=1}^{2}  \widehat{\mathcal{L}}^{\, \vec[b]}_{\vecs[k,a]} \right) \langle \zeta(\vecs [k,1]) \zeta(\vecs [k,2]) \rangle =  - \, 2 \bigg( \vec[b]\cdot\frac{\del}{\del \vecs[k,3]}\bigg) \frac{\langle \zeta(\vecs[k,1]) \zeta(\vecs[k,2]) \zeta(\vecs [k,3]) \rangle}{\langle \zeta(\vecs[k,3]) \zeta(-\vecs[k,3])\rangle'} \, \bigg|_{\vecs[k,3] \rightarrow 0},
  \end{equation}
  and
  \begin{equation}
  \label{limeq2}
  \left( \sum_{a=1}^{3}  \widehat{\mathcal{L}}^{\, \vec[b]}_{\vecs[k,a]} \right) \langle \zeta(\vecs [k,1]) \zeta(\vecs [k,2]) \zeta(\vecs [k,3]) \rangle =  - \, 2 \bigg( \vec[b]\cdot\frac{\del}{\del \vecs[k,4]}\bigg) \frac{\langle \zeta(\vecs[k,1]) \zeta(\vecs[k,2]) \zeta(\vecs[k,3]) \zeta(\vecs[k,4]) \rangle}{\langle \zeta(\vecs[k,4]) \zeta(-\vecs[k,4])\rangle'} \, \bigg|_{\vecs[k,4] \rightarrow 0}.
  \end{equation}
  For the special case $\vec[b] \propto \vecs[k,3]$ in eq.\eqref{limeq1} and $\vec[b] \propto \vecs[k,4]$ in eq.\eqref{limeq2}, we obtain the conformal consistency relations derived in eq.(37) of \cite{Creminelli:2012ed}. 
  However, as we will see in section \ref{wardcheck}, for the case $n\ge 4$ in eq.\eqref{wardt2} the extra terms in $\delta \zeta(\vec[k])$ due to the compensating spatial reparametrization cannot be neglected. Thus, the Ward identities obtained from eq.\eqref{wardt2} for $\vec[b] \propto \vecs[k,n+1]$ , $n\ge 4$, have additional terms as compared to eq.(37) of \cite{Creminelli:2012ed}.
  
 This concludes our summary of some of the main results of this paper.

 \section{Conditions on the Coefficient Functions}
 \label{conditions}
 In this section, we will show how the invariance of the wave function under the scale and special conformal transformations, eq. \eqref{scalechn} and eq.\eqref{sctn}, lead to conditions on the coefficient functions which are analogous to Ward identities in a conformal field theory. 
 These Ward identities also incorporate the breaking of conformal invariance in the inflationary background. 
 
 We start with the Ward identities of spatial reparametrizations in general. These will give rise to conditions analogous to Ward identities of stress-energy conservation in a field theory. We then obtain the identities for scale and special conformal transformations.
 
 In this section, it will be convenient to   work with a general metric perturbation ${\widehat \gamma}_{ij}$ without further imposing the transversality condition eq.\eqref{transverse}.
 Once the constraints on the coefficient functions have been obtained for these general perturbations, they will lead to constraints on expectation values which can be calculated only after  further gauge fixing, as explained in section \ref{identities} above.

\subsection{Spatial Reparametrizations}
\label{spreps}
Under a spatial reparametrization eq.\eqref{sprep}, the perturbations $\zeta$ and $\widehat\gamma_{ij}$ transforms as
\be
\zeta \rightarrow \zeta + \frac{1}{3}\, \del_i v_i + v^i \del_i \zeta  + \frac{1}{3} \, \del_i v_j \,\widehat\gamma_{ij} ,\label{zchgn}
\ee
and
\begin{equation} 
\begin{split}
\widehat{\gamma}_{ij} \rightarrow \widehat{\gamma}_{ij} &+ \bigg( \del_i v_j + \del_j v_i - \frac{2}{3} \, \del_a v_a \, \delta_{ij} \bigg) + \bigg(\widehat\gamma_{ik} \, \del_j v^k + \widehat\gamma_{jk} \, \del_i v^k \\ &+ v^k \del_k \widehat\gamma_{ij} - \frac{2}{3} \, \del_a v_a \, \widehat\gamma_{ij} - \frac{2}{3} \, \del_a v_b \, \widehat\gamma_{ab} \left(\delta_{ij} + \widehat\gamma_{ij}\right)\bigg).\label{hatgchn}
\end{split}
\end{equation}
See appendix \ref{perttrans} for some details of the derivation.

Now, for the invariance of the wave function under spatial reparametrizations, the  terms proportional to the transformation parameter $v_i$ that get generated because of the transformations eq.\eqref{zchgn}, eq.\eqref{hatgchn}, must cancel with one another. Consider first such terms  which are linear in $\zeta$. These  terms are produced by the first and second terms of the wave function, eq.\eqref{wfunction}. These are
\begin{equation}
\begin{split}
\label{lin1}
\delta\bigg( &- \half \int \mes[x] \, \mes[y] \, \zeta(\vec[x]) \zeta(\vec[y]) \, \langle T(\vec[x]) T(\vec[y]) \rangle \bigg) \\
&= - \frac{1}{6} \int \mes[x] \, \mes[y] \Big[ \del_i v_i(\vec[x])\, \zeta(\vec[y]) + \zeta(\vec[x]) \, \del_{y^{i}} v_i(\vec[y]) \Big] \langle T(\vec[x]) T(\vec[y]) \rangle,
\end{split}
\end{equation}
and
\begin{equation}
\begin{split}
\label{lin2}
\delta \bigg( - \int \mes[x] \, \mes[y] \, &\zeta(\vec[x]) \, \widehat\gamma_{ij}(\vec[y]) \, \langle T(\vec[x]) \widehat{T}^{ij}(\vec[y]) \rangle \bigg) \\ &= - \, 2 \int \mes[x] \, \mes[y] \, \zeta(\vec[x]) \,  \del_{y^{i}} v_j(\vec[y]) \, \langle T(\vec[x]) \widehat{T}^{ij}(\vec[y]) \rangle.
\end{split}
\end{equation}
Mutual cancellation of the terms in eq.\eqref{lin1} and eq.\eqref{lin2} produces the Ward identity\footnote{Note that unless otherwise stated, $\del_i$ stands for the derivative with respect to $x^i$, i.e. ${\del}/{\del x^i}$.}
\be
\label{linwa}
\del_i \langle \widehat{T}^{ij}(\vec[x]) T(\vec[y]) \rangle + \frac{1}{6} \, \del_j \langle T(\vec[x]) T(\vec[y]) \rangle = 0.
\ee
Using 
\be
\label{tdecom}
T^{ij} = \widehat{T}^{ij} + \frac{1}{3} \delta_{ij} \mathcal{T}, 
\ee
and eq.\eqref{relt}, this can also be written as 
\be
\label{wardl1}
\del_i \langle T^{ij}(\vec[x]) \mathcal{T}(\vec[y])\rangle = 0.
\ee
Similarly, canceling the extra terms in the wave function which are linear in $\widehat\gamma_{ij}$ gives
\be
\label{wardl2}
\del_i \langle T^{ij}(\vec[x]) {T}^{kl}(\vec[y])\rangle = 0.
\ee

Proceeding in a similar manner, and canceling terms linear in $v_i$ and quadratic in $\zeta$ gives
\begin{equation}
\begin{split}
\del_{i}\langle \widehat{T}^{ij}(\vec[x]) T(\vec[y]) T(\vec[z]) \rangle = \, & \half \, \del_j \big[\delta^3(\vec[x] - \vec[y])\big] \langle T(\vec[x]) T(\vec[z]) \rangle + \half \, \del_j \big[\delta^3(\vec[x] - \vec[z])\big] \langle T(\vec[x]) T(\vec[y]) \rangle \\ &- \frac{1}{6} \, \del_j\langle T(\vec[x]) \, T(\vec[y]) T(\vec[z]) \rangle . \label{wardq1}
\end{split}
\end{equation}
Eq.\eqref{wardq1} can be rewritten in terms of the complete stress-energy tensor $T^{ij}$ and its trace $\mathcal{T}$ as
\be
\label{wardq2}
\del_{i}\langle T^{ij}(\vec[x]) \mathcal{T}(\vec[y]) \mathcal{T}(\vec[z]) \rangle = \, \half \, \del_j \big[\delta^3(\vec[x] - \vec[y])\big] \langle \mathcal{T}(\vec[x]) \mathcal{T}(\vec[z]) \rangle + \half \, \del_j \big[\delta^3(\vec[x] - \vec[z])\big] \langle \mathcal{T}(\vec[x]) \mathcal{T}(\vec[y]) \rangle .
\ee

Similarly, canceling terms proportional to  $\zeta \, \widehat\gamma_{ij}$ produces the Ward identity
\begin{equation}
\begin{split}
\label{wardq3}
\del_{i} \langle T^{ij}(\vec[x]) \widehat{T}^{kl}(\vec[y]) T(\vec[z]) \rangle &= \half \, \del_j \big[\delta^3(\vec[x] - \vec[y])\big] \langle \widehat{T}^{kl}(\vec[x]) T(\vec[z])\rangle \\ &+ \half \, \del_j \big[\delta^3(\vec[x] - \vec[z])\big] \langle T(\vec[x]) \widehat{T}^{kl}(\vec[y])\rangle \\ &+ \frac{1}{3} \, \del_j \big[\delta^3(\vec[x] - \vec[y])\big] \langle \widehat{T}^{kl}(\vec[y]) T(\vec[z])\rangle \\ &- \delta_{jl} \, \del_i \big[\delta^3(\vec[x] - \vec[y])\big] \langle T^{ik}(\vec[y]) T(\vec[z])\rangle,
\end{split}
\end{equation}
and proportional to $\widehat\gamma_{ij} \widehat\gamma_{kl}$ gives

\begin{equation}
\begin{split}
\label{wardq4}
\del_{i} \langle T^{ij}(\vec[x]) \widehat{T}^{kl}(\vec[y]) \widehat{T}^{mn}(\vec[z]) \rangle &= \del_j \big[\delta^3(\vec[x] - \vec[z])\big] \langle \widehat{T}^{mn}(\vec[x]) \widehat{T}^{kl}(\vec[y]) \rangle \\
&+ \, \frac{2}{3} \, \del_j \big[\delta^3(\vec[x] - \vec[z])\big] \langle \widehat{T}^{kl}(\vec[y]) \widehat{T}^{mn}(\vec[z]) \rangle \\ &- 2 \, \delta_{jn} \, \del_i \big[\delta^3(\vec[x] - \vec[z])\big] \langle \widehat{T}^{kl}(\vec[y]) T^{im}(\vec[z]) \rangle.
\end{split}
\end{equation}

\subsection{Scale Transformations}
\label{wscale}

 We now turn to deriving conditions on the coefficient functions for the invariance of the wave function under scale transformations, eq.\eqref{scalechn}. The change in $\zeta$ and $\widehat\gamma_{ij}$ under these transformations is given by eq.\eqref{zchscal1}  and eq.\eqref{gchscal1} respectively.

 The procedure we follow is the same as outlined above. Canceling the  terms linear in the transformation parameter $\lambda$ and quadratic in $\zeta$ gives
\be
\label{wards1}
\int \mes[z] \, \langle T(\vec[x]) T(\vec[y]) T(\vec[z])\rangle = \left[ x^i \frac{\del}{\del x^i} + y^i \frac{\del}{\del y^i} \right] \langle T(\vec[x]) T(\vec[y]) \rangle + 6 \, \langle T(\vec[x]) T(\vec[y]) \rangle,
\ee
which in momentum space takes the form
\be
\label{wards2}
\lim_{\vecs[k,3] \rightarrow 0} \, \langle T(\vecs[k,1]) T(\vecs[k,2]) T(\vecs[k,3]) \rangle = - \left(  \sum_{a = 1}^2 k_a \, \frac{\del}{\del k_a} \right) \langle T(\vecs[k,1]) T(\vecs[k,2]) \rangle,
\ee
where $k_a \equiv |\vecs[k,a]|$. 

In general, the $n$-point correlation function of the $T$ operators will be related under scaling to the $(n-1)$-point correlation function through the relation
\be
\label{wards3}
\lim_{\vecs[k,n] \rightarrow 0} \, \langle T(\vecs[k,1]) \cdots T(\vecs[k,n]) \rangle = - \left(  \sum_{a = 1}^{n-1} k_a \, \frac{\del}{\del k_a} \right) \langle T(\vecs[k,1]) \cdots T(\vecs[k,n-1])\rangle.
\ee

Requiring the cancellation of the extra quadratic terms in $\widehat\gamma_{ij}$ gives us the Ward identity
\begin{equation}
\label{wards4}
\int \mes[z] \, \langle \widehat{T}^{ij}(\vec[x]) \widehat{T}^{kl}(\vec[y]) T(\vec[z])\rangle = \left[ x^i \frac{\del}{\del x^i} + y^i \frac{\del}{\del y^i} \right] \langle \widehat{T}^{ij}(\vec[x]) \widehat{T}^{kl}(\vec[y]) \rangle + 6 \, \langle \widehat{T}^{ij}(\vec[x]) \widehat{T}^{kl}(\vec[y]) \rangle ,
\end{equation}
which translates in momentum space to
\be
\label{wards5}
\lim_{\vecs[k,3] \rightarrow 0} \langle \widehat{T}^{ij}(\vecs[k,1]) \widehat{T}^{kl}(\vecs[k,2]) T(\vecs[k,3]) \rangle = - \left[  \sum_{a = 1}^2 k_a \, \frac{\del}{\del k_a} \right] \langle \widehat{T}^{ij}(\vecs[k,1]) \widehat{T}^{kl}(\vecs[k,2]) \rangle.
\ee

The general form of the scaling Ward identity relating the $n$-point correlation function of $(n-1)$ $\widehat{T}^{ij}$ operators and one insertion of $T$, to the $(n-1)$-point correlation function of the $\widehat{T}^{ij}$ operators is 

\begin{equation}
\begin{split}
\label{wards6}
\lim_{\vecs[k,n] \rightarrow 0} \langle \widehat{T}^{i_1 j_1}(\vecs[k,1]) \cdots &\widehat{T}^{i_{n-1} j_{n-1}}(\vecs[k,n-1]) T(\vecs[k,n]) \rangle =\\ &- \left[  \sum_{a = 1}^{n-1} k_a \, \frac{\del}{\del k_a} \right] \langle \widehat{T}^{i_1 j_1}(\vecs[k,1]) \cdots \widehat{T}^{i_{n-1} j_{n-1}}(\vecs[k,n-1]) \rangle.
\end{split}
\end{equation}

One can also write the general scaling Ward identity relating the $(n+1)$-point correlation function involving $m$ insertions of $\widehat{T}^{ij}$ and $(n+1-m)$ insertions of $T$, with the $n$-point correlation function of $m$ insertions of $\widehat{T}^{ij}$ and $(n-m)$ of $T$, as
\begin{equation}
\begin{split}
\label{wards7}
\lim_{\vecs[k,n+1] \rightarrow 0} \langle \widehat{T}^{i_1 j_1}(\vecs[k,1]) &\cdots \widehat{T}^{i_{m} j_m}(\vecs[k,m]) T(\vecs[k,m+1]) \cdots T(\vecs[k,n+1]) \rangle =\\ &- \left[  \sum_{a = 1}^{n} k_a \, \frac{\del}{\del k_a} \right] \langle \widehat{T}^{i_1 j_1}(\vecs[k,1]) \cdots \widehat{T}^{i_{m} j_m}(\vecs[k,m]) T(\vecs[k,m+1]) \cdots T(\vecs[k,n]) \rangle.
\end{split}
\end{equation}

One final comment. We began this subsection by considering terms which are quadratic in $\zeta$, eq.(\ref{wards1}). There is also a term which is linear in both $\zeta$ and the transformation parameter $\lambda$. Since $\lambda$ is spatially constant, this term has support only at zero momentum in the wave function and we  neglect it here.  However, in deriving expectation values, this term which is uncanceled  plays a crucial role, as was discussed in section \ref{scaletr} above. 

\subsection{Special Conformal Transformations}
\label{wsct}
We will now derive Ward identities for the invariance of the wave function under special conformal transformations, eq.\eqref{sctn}. The change in $\zeta$ and $\widehat\gamma_{ij}$ under this is given by eq.\eqref{sctzc} and eq.\eqref{sctgc}. Note that, as was mentioned at the beginning of this section,  we  are considering a general graviton perturbation here and have not fixed it  to be transverse; as a result we do not have to worry about the fact that a special conformal transformation leads to the gauge eq.\eqref{transverse} not being preserved.  

We start again with terms which are linear in $b_i$ and quadratic  in $\zeta$. Invariance of $\Psi$ then  gives
\begin{equation}
\begin{split}
\label{wardc1}
3 \, \vec[b] \cdot (\vec[x] + \vec[y])\, \langle T(\vec[x]) T(\vec[y]) \rangle &- \half \left[ \alpha^i(\vec[x]) \del_{x^i} + \alpha^i(\vec[y]) \del_{y^i} \right] \langle T(\vec[x]) T(\vec[y]) \rangle \\
&= \int \mes[z] \, (\vec[b]\cdot\vec[z]) \langle T(\vec[x]) T(\vec[y]) T(\vec[z])\rangle,
\end{split}
\end{equation}
which in momentum space has the form
\be
\label{wardc2}
\half \left[ \mathcal{L}^{\vec[b]}_{\vecs[k,1]} + \mathcal{L}^{\vec[b]}_{\vecs[k,2]} \right] \langle T(\vecs[k,1]) T(\vecs[k,2])\rangle = - \left( \vec[b] \cdot \frac{\del}{\del \vecs[k,3]} \right) \langle T(\vecs[k,1]) T(\vecs[k,2]) T(\vecs[k,3]) \rangle \bigg|_{\vecs[k,3] \rightarrow 0},
\ee
where $\mathcal{L}^{\vec[b]}_{\vec[k]}$ is the operator

\begin{equation}
\begin{split}
\label{opl}
\mathcal{L}^{\vec[b]}_{\vec[k]} &= 2 \, \Big( \vec[k] \cdot \frac{\del}{\del \vec[k]}\Big) \Big( \vec[b] \cdot \frac{\del}{\del \vec[k]} \Big) - (\vec[b]\cdot\vec[k]) \Big( \frac{\del}{\del \vec[k]} \cdot \frac{\del}{\del \vec[k]} \Big)
\\ &= (\vec[b] \cdot \vec[k]) \left( -\, \frac{2}{k} \frac{\del}{\del k} + \frac{\del^2}{\del k^2}\right).
\end{split}
\end{equation}
In general, the $n$-point correlation function of $T$ operators will be related to the $(n-1)$-point function under a special conformal transformation as
\be
\label{wardc3}
\half \left(\sum_{a=1}^{n-1} \mathcal{L}^{\vec[b]}_{\vecs[k,a]} \right) \langle T(\vecs[k,1]) \cdots T(\vecs[k,n-1])\rangle = - \bigg( \vec[b] \cdot \frac{\del}{\del \vecs[k,n]} \bigg) \langle T(\vecs[k,1]) \cdots T(\vecs[k,n])\rangle \bigg|_{\vecs[k,n] \rightarrow 0}.
\ee

Similarly, canceling the extra terms quadratic in $\widehat\gamma_{ij}$ gives
\begin{equation}
\label{wardc4}
\big[\mathcal{D}^{\vec[b]}_{\vec[x]} + \mathcal{D}^{\vec[b]}_{\vec[y]} \big] \langle \widehat{T}^{ij}(\vec[x]) \widehat{T}^{kl}(\vec[y]) \rangle = - \int \mes[z] \, (\vec[b] \cdot \vec[z]) \langle \widehat{T}^{ij}(\vec[x]) \widehat{T}^{kl}(\vec[y]) T(\vec[z])\rangle,
\end{equation}
where the action of the operator $\mathcal{D}^{\vec[b]}_{\vec[x]}$ is defined by
\begin{equation}
\begin{split}
\label{defax}
\mathcal{D}^{\vec[b]}_{\vec[x]} \, \widehat{T}^{ij}(\vec[x]) = - 3 \, (\vec[b]\cdot\vec[x]) \, \widehat{T}^{ij}(\vec[x]) &+ \half \, \alpha^m(\vec[x]) \, \del_m \widehat{T}^{ij}(\vec[x]) \\ &- \mathcal{M}^{\vec[b]}_{mi}(\vec[x]) \widehat{T}^{mj}(\vec[x]) - \mathcal{M}^{\vec[b]}_{mj}(\vec[x]) \widehat{T}^{im}(\vec[x]) ,
\end{split}
\end{equation}
with $\mathcal{M}_{ij}^{\vec[b]}(\vec[x])$ as defined in eq.\eqref{defmop}. The Ward identity eq. \eqref{wardc4} can be expressed in the momentum space as
\begin{equation}
\begin{split}
\label{wardc5}
\big[ \tilde{\mathcal{D}}^{\,\vec[b]}_{\vecs[k,1]} + \tilde{\mathcal{D}}^{\,\vec[b]}_{\vecs[k,2]} \big] \langle \widehat{T}^{ij}(\vecs[k,1]) \widehat{T}^{kl}(\vecs[k,2]) \rangle = - \left( \vec[b] \cdot \frac{\del}{\del \vecs[k,3]}\right)\langle \widehat{T}^{ij}(\vecs[k,1]) \widehat{T}^{kl}(\vecs[k,2]) T(\vecs[k,3])\rangle \bigg|_{\vecs[k,3] \rightarrow 0},
\end{split}
\end{equation}
where the momentum space operator $\tilde{\mathcal{D}}^{\,\vec[b]}_{\vec[k]}$ is defined by
\be
\label{defaxm}
\tilde{\mathcal{D}}^{\,\vec[b]}_{\vec[k]} \, \widehat{T}^{ij}(\vec[k]) = \half \, \mathcal{L}^{\vec[b]}_{\vec[k]} \, \widehat{T}^{ij}(\vec[k]) - \tilde{\mathcal{M}}^{\vec[b]}_{mi}(\vec[k]) \widehat{T}^{mj}(\vec[k]) - \tilde{\mathcal{M}}^{\vec[b]}_{mj}(\vec[k]) \widehat{T}^{im}(\vec[k]),
\ee
with the operator $\mathcal{L}^{\vec[b]}_{\vec[k]}$ as given in eq.\eqref{opl}, and $\tilde{\mathcal{M}}^{\vec[b]}_{ij}(\vec[k])$ as given by eq.\eqref{defmt}. 

Following a similar procedure as outlined above, we get the Ward identity for special conformal transformations relating the $n$-point correlation function with $(n-1)$ insertions of $\widehat{T}^{ij}$ and one insertion of $T$, to the $(n-1)$ point correlation function of $\widehat{T}^{ij}$ to be
\begin{equation}
\begin{split}
\label{wardc6}
\bigg( \sum_{a=1}^{n-1} \tilde{\mathcal{D}}^{\vec[b]}_{\vecs[k,a]} &\bigg) \langle \widehat{T}^{i_1 j_1}(\vecs[k,1]) \cdots \widehat{T}^{i_{n-1} j_{n-1}}(\vecs[k,n-1]) \rangle = \\ &- \bigg( \vec[b] \cdot \frac{\del}{\del \vecs[k,n]}\bigg) \langle \widehat{T}^{i_1 j_1}(\vecs[k,1]) \cdots \widehat{T}^{i_{n-1} j_{n-1}}(\vecs[k,n-1]) T(\vecs[k,n]) \rangle \bigg|_{\vecs[k,n] \rightarrow 0} .
\end{split}
\end{equation}

In general, the Ward identity of special conformal invariance relating the $n$-point correlation function with $m$ insertions of $\widehat{T}^{ij}$ and $(n-m)$ insertions of $T$, to the $(n+1)$-point correlation function with $m$ insertions of $\widehat{T}^{ij}$ and $(n+1-m)$ insertions of $T$ is given by
\begin{equation}
\begin{split}
\label{wardc7}
&\Bigg[\bigg(\sum_{a=1}^m \tilde{\mathcal{D}}^{\vec[b]}_{\vecs[k,a]} \bigg) + \half \bigg( \sum_{r=m+1}^{n} \mathcal{L}^{\vec[b]}_{\vecs[k,r]} \bigg)\Bigg] \langle \widehat{T}^{i_1 j_1}(\vecs[k,1]) \cdots \widehat{T}^{i_m j_m}(\vecs[k,m]) T(\vecs[k,m+1]) \cdots T(\vecs[k,n])\rangle \\ &= - \, \bigg( \vec[b] \cdot \frac{\del}{\del \vecs[k,n+1]} \bigg) \langle \widehat{T}^{i_1 j_1}(\vecs[k,1]) \cdots \widehat{T}^{i_m j_m}(\vecs[k,m]) T(\vecs[k,m+1]) \cdots T(\vecs[k,n+1])\rangle \bigg|_{\vecs[k,n+1] \rightarrow 0} ,
\end{split}
\end{equation}
where the operators $\tilde{\mathcal{D}}^{\vec[b]}_{\vec[k]}$ and $\mathcal{L}^{\vec[b]}_{\vec[k]}$ are given in eq.\eqref{defaxm} and eq.\eqref{opl} respectively. 

Finally, as in the case of the scale transformations, there is one remaining term linear in ${\widehat \gamma}_{ij}$ and $b_i$, with support at zero momentum, which is uncanceled. In calculating expectation values, section \ref{identities}, this term will vanish once we choose the transverse gauge, eq.\eqref{transverse}.

\section{Explicit Checks of the Special Conformal Ward Identities}
\label{wardcheck}
In this section, we present a few checks of the Ward identities of special conformal transformations, eq.\eqref{wardt2} and eq.\eqref{wardt3}. 
Our analysis above has been to the leading order in $H^2/M_{Pl}^2$, and we will verify the Ward identities to this order below.

\subsection{Basic Checks}
\label{bcwsct}
Consider the Ward identity eq.\eqref{wardt2} for the case of $n=2$. We have
\begin{equation}
\begin{split}
\label{check1}
&\bigg[ \sum_{a=1}^{2} \widehat{\mathcal{L}}^{\,\vec[b]}_{\vec[k_a]}\bigg] \langle \zeta(\vecs[k,1]) \zeta(\vecs[k,2]) \rangle + \bigg\{ \bigg(  6\, b^m k_1^i \int \frac{d^3\tilde{k}}{(2\pi)^3}\, \frac{1}{\tilde{k}^2}\,\langle \zeta(\vec[k_1] - \tilde{\vec[k]}) \zeta(\vecs[k,2]) \, \widehat\gamma_{im}(\tilde{\vec[k]}) \rangle
\\ &+ 2 \, b^m k_1^i \int \frac{d^3\tilde{k}}{(2\pi)^3}\, \frac{1}{\tilde{k}^2} \,\langle \widehat\gamma_{ij}(\vec[k_1] - \tilde{\vec[k]}) \, \widehat\gamma_{jm}(\tilde{\vec[k]})  \zeta(\vecs[k,2]) \rangle \bigg) + \Big( \vecs[k,1] \leftrightarrow \vecs[k,2]\Big) \bigg\}
\\ &= - \, 2 \bigg( \vec[b]\cdot\frac{\del}{\del \vecs[k,3]}\bigg) \frac{1}{\langle \zeta(\vecs[k,3]) \zeta(-\vecs[k,3])\rangle'} \, \langle \zeta(\vecs[k,1]) \zeta(\vecs[k,2]) \zeta(\vecs[k,3]) \rangle\bigg|_{\vecs[k,3] \rightarrow 0}.
\end{split}
\end{equation}

We will first consider the identity eq.\eqref{check1} in the de Sitter limit. We substitute
\be
\label{zdpr1}
\zeta = - \, \frac{H}{\dot{\bar{\phi}}} \, \delta\phi ,
\ee
and take the limit $\dot{\bar{\phi}} \rightarrow 0$. Keeping only the leading terms, we get
\begin{equation}
\begin{split}
\label{check2}
\bigg[\sum_{a=1}^{2} \widehat{\mathcal{L}}^{\,\vec[b]}_{\vec[k_a]}\bigg]\langle \delta \phi(\vecs[k,1]) \delta \phi(\vecs[k,2]) \rangle  = - \, \bigg\{\bigg(&6\, b^m k_1^i \int \frac{d^3\tilde{k}}{(2\pi)^3}\, \frac{1}{\tilde{k}^2}\,\langle \delta \phi(\vec[k_1] - \tilde{\vec[k]})  \delta \phi(\vecs[k,2]) \, \widehat\gamma_{im}(\tilde{\vec[k]}) \rangle \bigg) 
\\ &  + \Big( \vecs[k,1] \leftrightarrow \vecs[k,2] \Big) \bigg\}.
\end{split}
\end{equation}
Next, we introduce suitable factors of $H/M_{Pl}$. The wave function eq.\eqref{wfunction} arises from the action eq.\eqref{hdact}, which has a factor of $1/G \sim M_{Pl}^2$ in front of it. 
Thus, after suitably rescaling by powers of the Hubble parameter, $\Psi$ will go like 
\begin{equation} 
\begin{split}
\label{rewf1}
\Psi \sim \text{exp} \bigg[ &- \frac{M_{Pl}^2}{H^2}\bigg\lbrace\half \int \mes[x] \, \mes[y] \, \delta\phi(\vec[x]) \delta\phi(\vec[y]) \, \langle O(\vec[x]) O(\vec[y]) \rangle \\
&+ \half \int \mes[x]  \, \mes[y] \, \widehat\gamma_{ij}(\vec[x]) \widehat\gamma_{kl}(\vec[y]) \, \langle \widehat{T}^{ij}(\vec[x]) \widehat{T}^{kl}(\vec[y]) \rangle \\
&+ \frac{1}{3!} \int \mes[x] \, \mes[y] \, \mes[z] \, \delta\phi(\vec[x]) \delta\phi(\vec[y]) \delta\phi(\vec[z]) \, \langle O(\vec[x]) O(\vec[y]) O(\vec[z])\rangle \\
&+ \frac{1}{3!} \int \mes[x] \, \mes[y] \, \mes[z] \, \delta\phi(\vec[x]) \delta\phi(\vec[y]) \widehat\gamma_{ij}(\vec[z]) \, \langle O(\vec[x]) O(\vec[y]) \widehat{T}^{ij}(\vec[z])\rangle + \cdots \bigg\rbrace\bigg]. 
\end{split}
\end{equation}
As a result, we see that the propagators $\langle \delta\phi \, \delta\phi \rangle$ or $\langle \widehat\gamma_{ij} \widehat\gamma_{kl} \rangle$ behave like ${H^2/M_{Pl}^2}$, while each vertex, e.g., the three point vertices $\langle O(\vec[x]) O(\vec[y]) O(\vec[z])\rangle$ and $\langle O(\vec[x]) O(\vec[y]) \widehat{T}^{ij}(\vec[z])\rangle$ in eq.(\ref{rewf1}), go like $M_{Pl}^2/H^2$. With this, one can argue that 
\begin{equation}
\label{Mporder1}
 \begin{split}
  \langle\delta \phi \, \delta \phi \, \widehat\gamma_{ij} \rangle \sim \frac{H^4}{M^4_{Pl}}.
 \end{split}
\end{equation}

From eq.\eqref{Mporder1}, it becomes clear that to leading order in $H^2/M_{Pl}^2$, the correlation function $\langle \delta \phi(\vec[k_1] - \tilde{\vec[k]}) \delta \phi(\vecs[k,2]) \, \widehat\gamma_{im}(\tilde{\vec[k]}) \rangle$ in the RHS of eq.\eqref{check2} is suppressed compared to $\langle \delta \phi(\vecs[k,1]) \delta \phi(\vecs[k,2]) \rangle$ in the LHS and  eq.\eqref{check2} reduces to 
\be
\label{check3}
\bigg[\sum_{a=1}^{2} \widehat{\mathcal{L}}^{\,\vec[b]}_{\vec[k_a]}\bigg]\langle \delta \phi(\vecs[k,1]) \delta \phi(\vecs[k,2]) \rangle  = 0.
\ee
This condition is the statement of conformal invariance of the two point function $\langle \delta \phi(\vecs[k,1]) \delta \phi(\vecs[k,2]) \rangle$ in de Sitter space, and it is easy to verify that it is met. 

Next, let us consider the  $n=2$ scalar Ward identity, eq.\eqref{check1}, to the first non-trivial order in the slow roll approximation (but still to the leading order in $H^2/M_{Pl}^2$). 
 The terms proportional to $\langle \zeta \zeta \widehat\gamma\rangle$ and $\langle \widehat\gamma \widehat\gamma \zeta \rangle$ in eq.\eqref{check1} scale as $H^4/M_{Pl}^4$, and can be dropped compared to other terms, which scale as $H^2/M_{Pl}^2$, in the limit $H \ll M_{Pl}$. Eq.\eqref{check1} then reduces to
\begin{equation}
\begin{split}
\label{check5}
\bigg[ \sum_{a=1}^{2} \widehat{\mathcal{L}}^{\,\vec[b]}_{\vec[k_a]}\bigg] \langle \zeta(\vecs[k,1]) \zeta(\vecs[k,2]) \rangle = - \, 2 \bigg( \vec[b]\cdot\frac{\del}{\del \vecs[k,3]}\bigg) \frac{1}{\langle \zeta(\vecs[k,3]) \zeta(-\vecs[k,3])\rangle'} \, \langle \zeta(\vecs[k,1]) \zeta(\vecs[k,2]) \zeta(\vecs[k,3]) \rangle\bigg|_{\vecs[k,3] \rightarrow 0}.
\end{split}
\end{equation}
In the canonical slow-roll model we have 
\be
\label{zzdef1}
\langle \zeta(\vecs[k,1]) \zeta(\vecs[k,2]) \rangle = (2\pi)^3 \delta^3(\vecs[k,1]+\vecs[k,2])\, \frac{H^2}{M_{Pl}^2} \, \frac{1}{4 \epsilon} \, k_1^{\,-3 + n_S} ,
\ee
and\footnote{The result eq.\eqref{maldaeqt} is from \cite{Maldacena:2002vr}, with $k_t = k_1 + k_2 + k_3$.}
\begin{equation}
\begin{split}
\label{maldaeqt}
\langle \zeta(\vecs[k,1]) \zeta(\vecs[k,2]) \zeta(\vecs[k,3]) \rangle = (2\pi)^3 \delta^3&\Big(\sum_{a=1}^3\vecs[k,a]\Big) \frac{H^4}{M_{Pl}^4} \frac{1}{4\epsilon^2} \frac{1}{\prod_a (2 k_a)^3} \\ &\bigg[ (3\epsilon - 2\eta) \sum_{a=1}^3 k_a^{\,3} + 2\epsilon \, \bigg( \half \sum_{a\neq b} k_a k_b^{\,2} + \frac{4}{k_t} \sum_{a>b} k_a^{\,2} k_b^{\,2} \bigg)\bigg].
\end{split}
\end{equation}
Working to the leading order in the slow roll parameters $\epsilon, \eta$ as well as the scalar tilt $n_S$, one finds that eq.\eqref{check5} implies
\be
\label{checkr}
n_S = 2(\eta - 3\epsilon),
\ee
which is the correct result, \cite{Maldacena:2002vr}.

Note that the case $n=3$ for the Ward identity eq.\eqref{wardt2} in the slow roll approximation was discussed in detail in \cite{Kundu:2014gxa}.

Next, we turn to the graviton two-point correlator and consider the Ward identity eq.\eqref{wardt3}. Again dropping terms which are subleading in $H^2/M_{Pl}^2$, and working in the de Sitter limit $\dot{\bar{\phi}} \rightarrow 0$, we get,

\begin{equation}
\begin{split}
\label{check4}
\bigg[\sum_{a=1}^{2} \widehat{\mathcal{L}}^{\,\vec[b]}_{\vec[k_a]}\bigg]\langle \widehat\gamma_{ij}(\vecs[k,1])& \widehat\gamma_{kl}(\vecs[k,2])\rangle \\ = - &\Big( 2 \, \tilde{\mathcal{M}}_{im}^{\vec[b]} (\vecs[k,1]) \langle \widehat\gamma_{jm}(\vecs[k,1]) \widehat\gamma_{kl}(\vecs[k,2])\rangle + 2 \, \tilde{\mathcal{M}}_{jm}^{\vec[b]} (\vecs[k,1]) \langle \widehat\gamma_{im}(\vecs[k,1]) \widehat\gamma_{kl}(\vecs[k,2])\rangle \\
&+ 2 \, \tilde{\mathcal{M}}_{km}^{\vec[b]} (\vecs[k,2]) \langle \widehat\gamma_{ij}(\vecs[k,1]) \widehat\gamma_{ml}(\vecs[k,2])\rangle + 2 \, \tilde{\mathcal{M}}_{lm}^{\vec[b]} (\vecs[k,2]) \langle \widehat\gamma_{ij}(\vecs[k,1]) \widehat\gamma_{km}(\vecs[k,2])\rangle \\
&+ \frac{6 b^m }{k_1^2} \, \big(k_{1i} \, \langle \widehat\gamma_{jm}(\vecs[k,1]) \widehat\gamma_{kl}(\vecs[k,2])\rangle + k_{1j} \, \langle \widehat\gamma_{im}(\vecs[k,1]) \widehat\gamma_{kl}(\vecs[k,2])\rangle \big) \\
&+ \frac{6 b^m }{k_2^2} \, \big(k_{2k} \, \langle \widehat\gamma_{ij}(\vecs[k,1]) \widehat\gamma_{lm}(\vecs[k,2])\rangle + k_{2l} \, \langle \widehat\gamma_{ij}(\vecs[k,1]) \widehat\gamma_{km}(\vecs[k,2])\rangle \big) \Big).
\end{split}
\end{equation}
The two-point graviton correlator is
\begin{equation}
\begin{split}
\label{defp}
\langle \widehat\gamma_{ij}(\vec[k]) \widehat\gamma_{kl}(-\vec[k]) \rangle' = \frac{P_{ijkl}(\vec[k])}{k^3},
\end{split}
\end{equation}
where $P_{ijkl}(\vec[k])$ is given in eq.(5.2) of \cite{Ghosh:2014kba}. 
An explicit calculation then shows that eq.(\ref{check4})  is indeed met. 
Note that the last two terms on the RHS of  eq.(\ref{check4})  come from the compensating spatial reparametrization which maintains the transverse gauge for $\widehat\gamma_{ij}$.

\subsection{The Scalar Four Point Function}
\label{acwsct}
In this subsection, we consider  the Ward identity eq.\eqref{wardt2} for the case $n=4$. We will work to the leading order in $H^2/M_{Pl}^2$, and to the leading order in $\dot{\bar{\phi}}/H$, i.e. in the de Sitter limit. Eq.\eqref{wardt2} for the case of $n=4$ then gives
\begin{equation}
\begin{split}
\label{wdmomz1}
&\bigg[ \sum_{a=1}^{4} \widehat{\mathcal{L}}^{\,\vec[b]}_{\vec[k_a]}\bigg] \langle \zeta(\vecs[k,1]) \zeta(\vecs[k,2]) \zeta(\vecs[k,3])\zeta(\vecs[k,4]) \rangle + \bigg\{ \bigg(  2\, b^m k_1^i \int \frac{d^3\tilde{k}}{(2\pi)^3}\, \frac{1}{\tilde{k}^2} \times\\ & \Big[3\, \langle \zeta(\vec[k_1] - \tilde{\vec[k]}) \zeta(\vecs[k,2]) \zeta(\vecs[k,3])\zeta(\vecs[k,4]) \widehat\gamma_{im}(\tilde{\vec[k]})\rangle + \langle \widehat\gamma_{ij}(\vec[k_1] - \tilde{\vec[k]}) \, \widehat\gamma_{jm}(\tilde{\vec[k]})  \zeta(\vecs[k,2]) \zeta(\vecs[k,3])\zeta(\vecs[k,4]) \rangle \Big]\bigg) 
\\ &  + \big( \vecs[k,1] \leftrightarrow \vecs[k,2]\big)  + \big(\vecs[k,1] \leftrightarrow \vecs[k,3]\big) + \big(\vecs[k,1] \leftrightarrow \vecs[k,4]\big)\bigg\}
\\ &= - \, 2 \bigg( \vec[b]\cdot\frac{\del}{\del \vecs[k,5]}\bigg) \frac{1}{\langle \zeta(\vecs[k,5]) \zeta(-\vecs[k,5])\rangle'} \, \langle \zeta(\vecs[k,1]) \cdots \zeta(\vecs[k,5]) \rangle\bigg|_{\vecs[k,5] \rightarrow 0}.
\end{split}
\end{equation}
We next write eq.\eqref{wdmomz1} in terms of $\delta\phi$ by using eq.\eqref{zdpr1}, and take the de Sitter limit $\dot{\bar{\phi}} \rightarrow 0$. In this limit, the terms in eq.\eqref{wdmomz1} that survive are
\begin{equation}
\begin{split}
\label{wdmomz3}
&\bigg[\sum_{a=1}^{4} \widehat{\mathcal{L}}^{\,\vec[b]}_{\vec[k_a]}\bigg]\langle \delta \phi(\vecs[k,1]) \delta \phi(\vecs[k,2]) \delta \phi(\vecs[k,3])\delta \phi(\vecs[k,4]) \rangle  
\\ & = - \, \bigg\{\bigg(6\, b^m k_1^i \int \frac{d^3\tilde{k}}{(2\pi)^3}\, \frac{1}{\tilde{k}^2}\,\langle \delta \phi(\vec[k_1] - \tilde{\vec[k]}) \delta \phi(\vecs[k,2]) \delta \phi(\vecs[k,3]) \delta \phi(\vecs[k,4])  \widehat\gamma_{im}(\tilde{\vec[k]}) \rangle \bigg) 
\\ &  + \big( \vecs[k,1] \leftrightarrow \vecs[k,2] \big)  + \big( \vecs[k,1] \leftrightarrow \vecs[k,3] \big) + \big(\vecs[k,1] \leftrightarrow \vecs[k,4] \big)\bigg\}.
\end{split}
\end{equation}

Introducing suitable factors of $H/M_{Pl}$ by rescaling the wave function, eq.\eqref{rewf1}, we see that for connected correlators,
\begin{equation}
\label{Mplorder}
 \begin{split}
  \langle\delta \phi\delta \phi\delta \phi\delta \phi \rangle \sim \frac{H^6}{M^6_{Pl}} \, ; ~~
  \langle\delta \phi \widehat{\gamma}_{ij} \delta \phi\delta \phi\delta \phi \rangle \sim \frac{H^8}{M^8_{Pl}}.
 \end{split}
\end{equation}
From eq.\eqref{Mplorder}, it seems that for the Hubble scale being much small compared to the Planck scale, $H\ll M_{Pl}$, the correlation function $\langle \delta \phi(\vec[k_1] - \tilde{\vec[k]}) \delta \phi(\vecs[k,2]) \delta \phi(\vecs[k,3])\delta \phi(\vecs[k,4]) \widehat\gamma_{im}(\tilde{\vec[k]})\rangle$ in the RHS of eq.\eqref{wdmomz3} is suppressed compared to $\langle \delta \phi(\vecs[k,1]) \delta \phi(\vecs[k,2]) \delta \phi(\vecs[k,3])\delta \phi(\vecs[k,4]) \rangle$ in the LHS. However, one should also consider disconnected contributions to the RHS of eq.\eqref{wdmomz3} which may contribute to the same order of $H/M_{Pl}$ as the LHS. In particular, there is a disconnected contribution to the five point function $\langle\delta \phi\delta \phi\delta \phi\delta \phi \widehat{\gamma}_{ij} \rangle$ in eq.\eqref{wdmomz3}, which goes as
\begin{equation}
\label{Mplorder1}
  \langle\delta \phi \delta \phi \rangle \, \langle \delta \phi\delta \phi  \widehat{\gamma}_{ij} \rangle \sim \frac{H^6}{M^6_{Pl}},
\end{equation}
and which is of the same order as $\langle\delta \phi\delta \phi\delta \phi\delta \phi \rangle$.
With these considerations, eq.\eqref{wdmomz3} in the limit $H\ll M_{Pl}$ becomes
\begin{equation}
\begin{split}
\label{wdmomz4}
&\bigg[\sum_{a=1}^{4} \widehat{\mathcal{L}}^{\,\vec[b]}_{\vec[k_a]}\bigg]\langle \delta \phi(\vecs[k,1]) \delta \phi(\vecs[k,2]) \delta \phi(\vecs[k,3])\delta \phi(\vecs[k,4]) \rangle  
\\ & = - \, \bigg\{\bigg(6\, b^m k_1^i  \int \frac{d^3\tilde{k}}{(2\pi)^3}\, \frac{1}{\tilde{k}^2}\,\langle \delta \phi(\vec[k_1] - \tilde{\vec[k]}) \delta \phi(\vecs[k,2]) \rangle \langle \widehat\gamma_{im}(\tilde{\vec[k]})  \delta \phi(\vecs[k,3])\delta \phi(\vecs[k,4]) \rangle 
\\ & + \big( \vecs[k,2] \leftrightarrow \vecs[k,3] \big)  + \big( \vecs[k,2] \leftrightarrow \vecs[k,4]\big) \bigg) 
 + \big( \vecs[k,1] \leftrightarrow \vecs[k,2]\big)  + \big(\vecs[k,1] \leftrightarrow \vecs[k,3]\big) + \big(\vecs[k,1] \leftrightarrow \vecs[k,4]\big)\bigg\}.
\end{split}
\end{equation}

It is important to note that there are other possible disconnected contributions to the five point correlator $\langle \delta \phi(\vec[k_1] - \tilde{\vec[k]}) \delta \phi(\vecs[k,2]) \delta \phi(\vecs[k,3])\delta \phi(\vecs[k,4]) \widehat\gamma_{im}(\tilde{\vec[k]}) \rangle$, such as 
\begin{equation}
 \begin{split}
 \label{disn}
 \langle  \delta \phi(\vecs[k,2])\delta \phi(\vecs[k,3]) \rangle \langle \widehat\gamma_{im}(\tilde{\vec[k]}) \delta \phi(\vec[k_1] - \tilde{\vec[k]}) \delta \phi(\vecs[k,4]) \rangle .
 \end{split}
\end{equation}
However, this requires $\vecs[k,2] + \vecs[k,3] = \vecs[k,1] + \vecs[k,4] = 0$, and will not contribute in general. 

Eq.\eqref{wdmomz4} gives the change in the four point correlator $\langle \delta\phi \delta\phi \delta\phi \delta\phi \rangle$ under a special conformal transformation in the exact de Sitter limit.
Using the relations
\begin{equation}
\begin{split} \label{rel1}
\langle  \delta \phi(\vecs[k,1])\delta \phi(\vecs[k,2]) \rangle& = (2\pi)^3 \delta^3(\vecs[k,1] + \vecs[k,2]) \langle  \delta \phi(\vecs[k,1])\delta \phi(\vecs[k,2]) \rangle'
\\& ={(2\pi)^3 \delta^3(\vecs[k,1] + \vecs[k,2])} \, \frac{H^2}{M_{Pl}^2} \frac{1}{2 k_1^3},
\end{split}
\end{equation}
and
\begin{equation}
\begin{split} \label{rel2}
\langle \widehat\gamma_{im}(\tilde{\vec[k]})  \delta \phi(\vecs[k,3])\delta \phi(\vecs[k,4]) \rangle = & -2 \,\langle  \delta \phi(\vecs[k,3])\delta \phi(-\vecs[k,3]) \rangle' \langle  \delta \phi(\vecs[k,4])\delta \phi(-\vecs[k,4]) \rangle'
\langle \widehat\gamma_{im}(\tilde{\vec[k]}) \widehat\gamma_{kl}(-\tilde{\vec[k]}) \rangle' \\ &  \hspace{8mm} \langle\widehat T_{kl}(\tilde{\vec[k]})  O(\vecs[k,3]) O(\vecs[k,4]) \rangle
\\ =& - 2\, \frac{H^4}{M_{Pl}^4} {\langle \widehat\gamma_{im}(\tilde{\vec[k]}) \widehat\gamma_{kl}(-\tilde{\vec[k]}) \rangle' \, \langle\widehat T_{kl}(\tilde{\vec[k]})  O(\vecs[k,3]) O(\vecs[k,4]) \rangle \over (2k_3^3)(2k_4^3)} ,
\end{split}
\end{equation}
we can write eq.\eqref{wdmomz4} as
\begin{equation}
\begin{split}
\label{wdmomz6}
\bigg[\sum_{a=1}^{4} \widehat{\mathcal{L}}^{\,\vec[b]}_{\vec[k_a]}\bigg]\langle \delta \phi(\vecs[k,1]) \delta \phi(\vecs[k,2])  \delta \phi(\vecs[k,3])\delta \phi(\vecs[k,4]) \rangle &= \\ \bigg[\frac{H^6}{M_{Pl}^6} {6 b^m P_{imkl}(\vec[k_1]+\vec[k_2]) \over |\vec[k_1]+\vec[k_2]|^5} \bigg\{ \bigg({k_{1i}\over k_{2}^3}&+{k_{2i}\over k_{1}^3} \bigg) {\langle\widehat T_{kl}(\vec[k_1]+\vec[k_2]) O(\vecs[k,3]) O(\vecs[k,4]) \rangle \over (2k_3^3)(2k_4^3)} \\ + \bigg({k_{3i}\over k_{4}^3}&+{k_{4i}\over k_{3}^3} \bigg) {\langle\widehat T_{kl}(\vec[k_3]+\vec[k_4])  O(\vecs[k,1]) O(\vecs[k,2]) \rangle \over (2k_1^3)(2k_2^3)} \bigg\} \bigg]
\\ +\Big[\vec[k_2]& \leftrightarrow \vec[k_3]\Big]+ \Big[\vec[k_2] \leftrightarrow \vec[k_4]\Big],
\end{split}
\end{equation}
where we have used the eq.\eqref{defp}.

Eq.\eqref{wdmomz6} gives us the change in the scalar four point function under a special conformal transformation. We can verify this by performing an explicit check. The four point function $\langle \delta\phi \delta\phi \delta\phi \delta\phi\rangle$ was calculated in \cite{Ghosh:2014kba}. It is given by
\be
\label{def4pt}
\langle \delta \phi(\vecs[k,1]) \delta \phi(\vecs[k,2]) \delta \phi(\vecs[k,3])\delta \phi(\vecs[k,4]) \rangle = \int [\mathcal{D}\delta\phi] \, \delta \phi(\vecs[k,1]) \delta \phi(\vecs[k,2]) \delta \phi(\vecs[k,3])\delta \phi(\vecs[k,4]) P[\delta\phi],
\ee
where $P[\delta\phi]$ is the probability distribution function
\be
\label{pdp2}
P[\delta\phi] = \int [\mathcal{D}\widehat{\gamma}_{ij}] \, \big|\Psi[\delta\phi, \widehat{\gamma}_{ij}]\big|^2.
\ee
An explicit expression for $P[\delta\phi]$ was obtained starting from the wave function, eq.\eqref{rewf1}, in eq.(5.3) of \cite{Ghosh:2014kba}. It is given by\footnote{In \cite{Ghosh:2014kba}, the term $\langle\widehat T_{kl} OO\rangle$ in the wave function appeared with a coefficient $1/4$ (see eq(2.36) of \cite{Ghosh:2014kba}). But in the present work we choose to have a $1/2$, which means we need to consistently replace  
\begin{equation*}
\langle\widehat T_{kl} OO\rangle_{there} \rightarrow 2 \langle\widehat T_{kl} OO\rangle_{here} 
\end{equation*}
while using expressions from \cite{Ghosh:2014kba}.}
\begin{equation}
\label{valpphi1}
 \begin{split}
  P[\delta \phi] =  \exp \bigg[& {M_{Pl}^2 \over H^2} \bigg(-\int {d^3k_1 \over (2\pi)^3} {d^3k_2 \over (2\pi)^3} \ \delta \phi(\vk) \delta \phi(\vkk) \, \langle O(-\vk) O(-\vkk)\rangle  \\ 
  &+ \int \prod_{J=1}^4 \bigg\{{d^3k_J \over (2\pi)^3}\delta \phi({\vect{k}}_J)\bigg\}
  \bigg\{- {1 \over 12} \langle O(-\vk) O(-\vkk)O(-\vkkk) O(-\vkfour)\rangle \\
&+ {1 \over 2} \langle O(-\vk) O(-\vkk) \widehat{T}_{ij}(\vk+\vkk)\rangle' \, \langle O(-\vkkk) O(-\vkfour) \widehat{T}_{kl}(\vkkk+\vkfour)\rangle'  \\ & \hspace{5mm} (2 \pi)^3  \delta^3 \bigg(\sum_{J=1}^4 {\vect{k}}_J \bigg) P_{i j k l}(\vk+\vkk) {1\over |\vk+\vkk|^3} \bigg\} \bigg) \bigg].
 \end{split}
\end{equation}
From eq.\eqref{def4pt} and eq.\eqref{valpphi1}, we see that the four point function has two types of contributions,
\be
\label{decom4}
\langle \delta\phi \delta\phi \delta\phi \delta\phi\rangle = \langle \delta\phi \delta\phi \delta\phi \delta\phi\rangle_{CF} + \langle \delta\phi \delta\phi \delta\phi \delta\phi\rangle_{ET} .
\ee
Here, $\langle \delta\phi \delta\phi \delta\phi \delta\phi\rangle_{CF}$ is the term proportional to $\langle OOOO \rangle$,
\be
\label{defcf}
\langle \delta\phi(\vecs[k,1]) \delta\phi(\vecs[k,2]) \delta\phi(\vecs[k,3]) \delta\phi(\vecs[k,4])\rangle_{CF} = - \frac{1}{8} \frac{H^6}{M_{Pl}^6} \frac{1}{\prod_{a=1}^{4} k_a^{\,3}} \langle O(\vecs[k,1])O(\vecs[k,2])O(\vecs[k,3])O(\vecs[k,4]) \rangle,
\ee
and $\langle \delta\phi \delta\phi \delta\phi \delta\phi\rangle_{ET}$ is the term proportional to $\langle OO\widehat{T}_{ij} \rangle' \langle OO\widehat{T}_{kl} \rangle'$,
\begin{equation}
\begin{split}
\label{dp4et}
\langle \delta\phi(\vecs[k,1]) \delta\phi(\vecs[k,2]) \delta\phi(\vecs[k,3]) \delta\phi(\vecs[k,4])\rangle_{ET} &= \frac{1}{4} \frac{H^6}{M_{Pl}^6} \frac{1}{\prod_{a=1}^{4} k_a^{\,3}} \, \Big[ I(\vecs[k,1], \vecs[k,2], \vecs[k,3], \vecs[k,4]) \\ + I(\vecs[k,1], \vecs[k,3], \vecs[k,2], \vecs[k,4]) &+ I(\vecs[k,1], \vecs[k,4], \vecs[k,3], \vecs[k,2]) + I(\vecs[k,3], \vecs[k,2], \vecs[k,1], \vecs[k,4]) \\ &+ I(\vecs[k,4], \vecs[k,2], \vecs[k,3], \vecs[k,1]) + I(\vecs[k,3], \vecs[k,4], \vecs[k,1], \vecs[k,2]) \Big],
\end{split}
\end{equation}
where $I(\vecs[k,1], \vecs[k,2], \vecs[k,3], \vecs[k,4])$ is given in eq.(E.13) of \cite{Ghosh:2014kba},
\begin{equation}
\begin{split}
\label{iexp}
I(\vecs[k,1], \vecs[k,2], \vecs[k,3], \vecs[k,4]) = \int {d^3k_5 \over (2\pi)^3}{d^3k_6 \over (2\pi)^3} &{\langle \widehat{T}_{ij} ({\vect{k}}_5) \widehat{T}_{kl}({\vect{k}}_6)\rangle \over k_5^3  \ k_6^3} \\ &\langle O(\vk)O(\vkk) \widehat{T}_{ij}({\vect{k}}_5)\rangle \langle O(\vkkk)O(\vkfour) \widehat{T}_{kl}({\vect{k}}_6)\rangle.
\end{split}
\end{equation}

Now, the term $\langle \delta\phi \delta\phi \delta\phi \delta\phi\rangle_{CF}$ is invariant under a special conformal transformation, whereas the term $\langle \delta\phi \delta\phi \delta\phi \delta\phi\rangle_{ET}$ does change. We therefore have
\be
\label{equal}
\bigg[\sum_{a=1}^{4} \widehat{\mathcal{L}}^{\,\vec[b]}_{\vec[k_a]}\bigg]\langle \delta \phi(\vecs[k,1]) \delta \phi(\vecs[k,2]) \delta \phi(\vecs[k,3]) \delta \phi(\vecs[k,4]) \rangle = \bigg[\sum_{a=1}^{4} \widehat{\mathcal{L}}^{\,\vec[b]}_{\vec[k_a]}\bigg]\langle \delta \phi(\vecs[k,1]) \delta \phi(\vecs[k,2]) \delta \phi(\vecs[k,3]) \delta \phi(\vecs[k,4]) \rangle_{ET} .
\ee
As discussed in appendix (E.2) of \cite{Ghosh:2014kba}, we have
\begin{equation}
\begin{split}
\label{dpsc4t}
\bigg[\sum_{a=1}^{4} \widehat{\mathcal{L}}^{\,\vec[b]}_{\vec[k_a]}\bigg]\langle \delta \phi(\vecs[k,1]) \delta \phi(\vecs[k,2]) \delta \phi(\vecs[k,3]) \delta \phi(&\vecs[k,4]) \rangle_{ET} = \frac{1}{4} \frac{H^6}{M_{Pl}^6} \frac{1}{\prod_{a=1}^{4} k_a^{\,3}} \, \Big[ \delta^C I(\vecs[k,1], \vecs[k,2], \vecs[k,3], \vecs[k,4]) \\ + \delta^C I(\vecs[k,1], \vecs[k,3], \vecs[k,2], \vecs[k,4]) &+ \delta^C I(\vecs[k,1], \vecs[k,4], \vecs[k,3], \vecs[k,2]) + \delta^C I(\vecs[k,3], \vecs[k,2], \vecs[k,1], \vecs[k,4]) \\ &+ \delta^C I(\vecs[k,4], \vecs[k,2], \vecs[k,3], \vecs[k,1]) + \delta^C I(\vecs[k,3], \vecs[k,4], \vecs[k,1], \vecs[k,2]) \Big],
\end{split}
\end{equation}
with $\delta^C I(\vecs[k,1], \vecs[k,2], \vecs[k,3], \vecs[k,4])$ given in eq.(E.23) of \cite{Ghosh:2014kba},
\begin{equation}
\begin{split}
\label{e23}
 \delta^C I(\vecs[k,1], \vecs[k,2], \vecs[k,3], \vecs[k,4]) = 12 b_m \int {d^3k \over (2\pi)^3} {P_{imkl}( {\vect{k}}) \over  k^5} \, k_{j} \, \langle  O(\vk)&O(\vkk)\widehat{T}_{ij}({\vect{k}})\rangle  \\ &\langle O(\vkkk) O(\vkfour) \widehat{T}_{kl}(-{\vect{k}})\rangle.
 \end{split}
\end{equation}
By using the Ward identity eq.(3.8) of \cite{Ghosh:2014kba} expressed in momentum space,
\begin{equation}
\label{wdk}
k_j \langle\widehat T_{ij}(\vec[k]) O(\vec[k'_1]) O(\vec[k'_2]) \rangle  = {1 \over 2}\Big(k'_{2i} \, \langle O(\vec[k'_1]+\vec[k]) O(\vec[k'_2]) \rangle + k'_{1i} \, \langle O(\vec[k'_2]+\vec[k]) O(\vec[k'_1])\rangle\Big) , 
\end{equation}
we can calculate the RHS of eq.\eqref{dpsc4t}. This gives the result eq.\eqref{wdmomz6} for the change in the four point function, and completes the check.

\section{Late Time Behaviour of Modes}
\label{latebh}
In this section, we elaborate on the late time behaviour of modes in the canonical model of slow roll inflation and also after including higher derivative terms in the action. 

\subsection{The Canonical Model of Slow Roll Inflation}
\label{canmodel}
We have discussed in section \ref{basics} that one can use the residual time reparametrization invariance, eq.\eqref{timere}, in the gauge eq.\eqref{gfix}, to set $\delta \phi=0$, and that the remaining perturbations all become time independent in this gauge, at late times. 
Here we demonstrate this behaviour explicitly in the  canonical slow roll model of inflation. The behaviour in the presence of higher derivative terms is discussed in section \ref{hdcorrect}.

The action for the  canonical model of slow roll inflation is given by
\be
\label{canact}
S = M_{Pl}^2 \int d^4x \sqrt{-g} \left( \half R - \half (\nabla\phi)^2 - V(\phi) \right).
\ee
In the canonical model, the Hubble parameter eq.\eqref{hubble} is given by
\be
\label{hubb}
H^2 = \frac{V}{3}.
\ee
The background $\bar\phi(t)$ satisfies the equation of motion
\be
\label{bck}
\ddot{\bar{\phi}} + 3 H \dot{\bar{\phi}} + V'(\bar{\phi}) = 0 ,
\ee
which in the slow roll approximation reduces to
\be
\label{backe}
\dot{\bar{\phi}} \approx - \frac{V'}{3H},
\ee
where a $'$ denotes a derivative with respect to the scalar field. Using eq.\eqref{hubb} and eq.\eqref{backe}, the slow roll parameters $\epsilon_1, \delta$ and $\epsilon$, defined in eq.\eqref{epsdef}, eq.\eqref{deltadef} and eq.\eqref{eps1}, can be expressed as
\be
\label{srparsr}
\epsilon_1 = \epsilon = \half \left( \frac{V'}{V} \right)^2, \, \delta = \epsilon_1 - \frac{V''}{V}.
\ee
The slow roll conditions, eq.\eqref{slrcond}, are 
\be
\label{scondv}
\left( \frac{V'}{V} \right)^2 \ll 1, \, \, \frac{V''}{V} \ll 1.
\ee

For the purpose of convenience in calculations, it is helpful to further decompose the metric perturbation $\gamma_{ij}$, eq.\eqref{pert_met}, as follows \cite{Weinberg:2008zzc}    
\be
\label{pertmd}
\gamma_{ij} =  [A \, \delta_{ij} + \del_i \del_j B + \del_i C_j + \del_j C_i + D_{ij}],
\ee
where $A, B$ transform as scalars, $C_i$ transforms like a 3-vector and $D_{ij}$ transforms as a rank-2 tensor under spatial rotations. Note that the perturbation $C_i$ is divergence-less, and $D_{ij}$ is transverse and traceless,
\be
\label{dttt}
\del_i C_i = 0, \, \del_i D_{ij} = 0, \, \, D_{ii} = 0.
\ee

The Einstein equations to linear order in the perturbations about the FRW inflationary background are given by (see appendix \ref{deteins})\footnote{For brevity, we present the equations in units with $M_{Pl}^2 = \frac{1}{8\pi G} = 1$.}
\be
\label{eqa}
- \, V'(\bar{\phi}) \, \delta\phi = \frac{1}{2a^2} \nabla^2 A - \half \ddot{A} -3 \left( \frac{\dot{a}}{a}\right) \dot{A} - \half \left( \frac{\dot{a}}{a}\right) \nabla^2 \dot{B},
\ee
\be
\label{eqb}
0 = A - a^2 \ddot{B} - 3 a \dot{a} \dot{B} ,
\ee
\be
\label{eqc}
- \, \dot{\bar{\phi}} \, \delta\phi = \dot{A} ,
\ee
\be
\label{eqd}
- \, \left(2 \, \dot{\bar{\phi}} \, \delta\dot{\phi} - \, V'(\bar{\phi}) \, \delta\phi \right) = \frac{3}{2} \ddot{A} + 3 \left( \frac{\dot{a}}{a}\right) \dot{A} + \half \nabla^2 \ddot{B} + \left( \frac{\dot{a}}{a}\right) \nabla^2 \dot{B} ,
\ee
for the scalar perturbations $A, B, \delta \phi$. The vector perturbations $C_i$ satisfy the equation
\be
\label{eqe}
\nabla^2 \dot{C}_i = 0.
\ee
For the tensor perturbations $D_{ij}$ we get
\be
\label{eqf}
\ddot{D}_{ij} + 3 \left( \frac{\dot{a}}{a}\right) \dot{D}_{ij} - \frac{1}{a^2} \nabla^2 D_{ij} = 0.
\ee
The equation of motion for $\delta \phi $ is
\be
\delta\ddot{\phi} + 3 \left( \frac{\dot{a}}{a}\right) \delta\dot{\phi} + V''(\bar{\phi}) \, \delta\phi - \frac{1}{a^2} \nabla^2 \delta\phi = - \half \, \dot{\bar{\phi}} \left( 3 \dot{A} + \nabla^2 \dot{B} \right). \label{eomdp}
\ee

Eq.\eqref{eqb} can be used to solve for $B$ in terms of $A$, 
\be
\label{gsolb}
B(t,\vec[x]) = \int^{\,t} dt' \, \frac{1}{a(t')^3} \left( G_1(\vec[x]) + \int^{\,t'} dt'' \, a(t'') A(t'',\vec[x]) \right) + G_2(\vec[x]) ,
\ee
where $G_1, G_2$ are arbitrary functions of $\vec[x]$. 

The late time behaviour of these equations can be obtained by dropping all spatial derivatives of the form $\nabla^2/a^2$ in eqs.\eqref{eqa}, \eqref{eqf} and \eqref{eomdp}. In addition, due to the $1/a^3$ pre-factor in eq.\eqref{gsolb}, we get that 
\be
\label{blsol}
B(t,\vec[x]) \approx G_2(\vec[x]) \; \text{for t} \rightarrow \infty,
\ee
so that all time derivatives of $B$ vanish at late times. Equations \eqref{eqa}, \eqref{eqd} and \eqref{eqf} then simplify to
\be
\label{eqg}
V'(\bar{\phi}) \, \delta\phi = \half \ddot{A} + 3 \left( \frac{\dot{a}}{a}\right) \dot{A} ,
\ee
\be
\label{eqh}
- \, \left(2 \, \dot{\bar{\phi}} \, \delta\dot{\phi} - \, V'(\bar{\phi}) \, \delta\phi \right) = \frac{3}{2} \ddot{A} + 3 \left( \frac{\dot{a}}{a}\right) \dot{A} ,
\ee
\be
\label{eqi}
\ddot{D}_{ij} + 3 \left( \frac{\dot{a}}{a}\right) \dot{D}_{ij} = 0,
\ee
and eq.\eqref{eomdp} becomes
\be
\delta\ddot{\phi} + 3 \left( \frac{\dot{a}}{a}\right) \delta\dot{\phi} + V''(\bar{\phi}) \, \delta\phi + \frac{3}{2} \, \dot{\bar{\phi}} \dot{A} = 0. \label{eqj}
\ee

 As discussed in appendix \ref{deteins}, the late time behaviour for $A, \delta \phi$ is 
 \be
\label{gsola}
A(t,\vec[x]) = P_1(\vec[x]) - 2 \left( \frac{\dot{a}}{a}\right) P_2(\vec[x]),
\ee
and
\be
\label{gsoldp}
\delta\phi(t,\vec[x]) = -\, \dot{\bar{\phi}}(t) \, P_2(\vec[x]),
\ee
 where  $P_1, P_2$ are time independent functions of $\vec[x]$. Also, the perturbations $C_i, D_{ij}$ become time independent.

We can now carry out a time reparametrization 
\be
\label{tren}
t \rightarrow t + P_2(\vec[x]),
\ee
along with the accompanying spatial reparametrization, eq.\eqref{xrep}, which maintains the gauge choice eq.\eqref{gfix}. Note that under the time reparametrization eq.\eqref{timere}, and the accompanying spatial reparametrization eq.\eqref{xrep}, the perturbations change as

\be
\delta A = 2 \left( \frac{\dot{a}}{a}\right) \epsilon(\vec[x]) , \label{chnga} 
\ee
\be
\delta B = 2 \, \epsilon(\vec[x]) \int^t dt' \, \frac{1}{a^2(t')} , \label{chngb} 
\ee
\be
\delta C_i = 0, \label{chngc} 
\ee
\be
\delta D_{ij} = 0 , \label{chngd} 
\ee
\be
\delta (\delta\phi) = \dot{\bar{\phi}} \, \epsilon(\vec[x]). \label{chngdp}
\ee
We see from eq.\eqref{gsoldp} and eq.\eqref{chngdp} that the change eq.\eqref{tren} sets the late time value of $\delta \phi$ to vanish. 
In addition, using eq.\eqref{chnga} we see that the value of $A$ is given by 
\be
\label{lvala}
A \rightarrow A' = P_1(\vec[x]) ,
\ee
while $C$, $D_{ij}$ are unchanged and therefore continue to be time independent. 
$B$ is changed by the the time reparametrization eq.\eqref{tren}, see eq.\eqref{chngb}, but this change vanishes at late times, and thus $B$ too continues to be time independent. 
Thus, we see that in the gauge $\delta \phi=0$ all the perturbations freeze out at late times \footnote{The perturbations $\zeta, {\widehat \gamma}_{ij}$, which appear in section \ref{basics} and the discussion thereafter, are given by
\begin{equation*}
\zeta = \half A + \frac{1}{6} \nabla^2 B\, , \;\;\; \text{and}\;\;\;
\widehat{\gamma}_{ij} = D_{ij} + \del_i C_j + \del_j C_i + \del_i \del_j B - \frac{1}{3}\, \delta_{ij} \nabla^2 B.
\end{equation*}
}. 

\subsection{Higher Derivative Corrections}
\label{hdcorrect}
In the discussion above, we have considered the canonical model of slow roll inflation, with the action in eq.\eqref{canact}. 
The action for this model involves two-derivative terms. One of the main motivations of our work is to be able to use symmetry considerations in more complicated situations where explicit computations or models may be unavailable.
An example is the possibility that the Hubble scale $H$ during inflation is of order the string scale $M_{st}$, so that  higher derivative corrections to eq.\eqref{canact} would be important.  Given our limited  knowledge  of string theory in time dependent situations,
explicit models or calculations for such a scenario are not possible today.  But a  symmetry based analysis should still be possible, as we discuss further in this section. 

The more general situation we have in mind is the one with an effective action having higher order terms of the schematic form
\be
\label{hdact}
S = \frac{1}{16 \pi G} \int d^4x \sqrt{-g} \left[R + (\partial \phi)^2 - 2 V + {R^2 \over  \Lambda^2} + {R^3 \over \Lambda^4} + { (\partial \phi)^4\over \Lambda^2} + \cdots \right].
\ee
The higher derivative terms are important because $H \sim O(\Lambda)$. In eq.\eqref{hdact}, $\Lambda$ is the underlying cutoff scale, which would be of order the string scale $M_{st}$ in string theory. The term  ${R^2/ \Lambda^2}$ schematically denotes four derivative terms, and so on. 
Also, the coefficients of each of the higher order terms can in general  be a function of $\phi$. Let us  note that in the background solution the  contribution from terms like $(\partial \phi)^4$ will be small, since the inflaton will be evolving slowly.  However, these terms will be important in determining the behaviour of the perturbations, since the perturbations will start out with  physical wavelengths $\lambda \ll H^{-1}$, and then freeze out at a time when $\lambda \simeq H^{-1}$. 

Let us   note that $H \sim O(\Lambda)$ is consistent with the bounds on the tensor perturbations, since $\Lambda$ can be much smaller than $M_{pl}$, as indeed happens in weakly coupled string theory. The condition $\Lambda \ll M_{Pl}$ also  ensures that all quantum loop effects are small, and it is only tree level effects involving  the higher derivative corrections which are important in the kind of scenario we have in mind. 

In fact,  considerations of the last few sections can be extended in a straightforward way to  situations of this  type. The crucial point is that even with the higher derivative terms present, one can argue that solutions with the same asymptotic behaviour as in the two-derivative case continue to  exist. The underlying reason for this is that the asymptotic behaviour in the two-derivative case follows from gauge invariance. We will discuss this in more detail in the next subsection. 
Given this fact, the arguments leading to the Ward identities can be easily seen to apply in cases with higher derivative corrections as well. A further change of coordinates allows us to set $\delta \phi$ to vanish, as discussed in section \ref{basics}, and the invariance of the wave function under the residual spatial reparametrizations in the synchronous gauge then leads to the Ward identities of interest. 

\subsubsection{Freezing of the Perturbations}
\label{pertfreeze}
We start with a discussion of the spin-$2$ component $D_{ij}$, eq.\eqref{pertmd}, which corresponds to gravity waves. 
In the two-derivative theory it satisfies the eq.\eqref{eqf}. At late times, when $k^2/a^2$ becomes sufficiently small, this becomes eq.\eqref{eqi}, which has the general solution eq.\eqref{gsold}. In particular, $D_{ij}$ becomes time independent, satisfying eq.\eqref{llimd}, since the additional solution proportional to $K_{ij}$ in eq.\eqref{gsold} dies out as $t \rightarrow \infty$. 
Higher  derivative terms   would result in contributions  to  the equation of motion with  either additional spatial derivatives, and/or  additional time derivatives. All terms with spatial derivatives will become small, since the physical wavelength $\lambda$ for fixed $\vec[k]$ becomes large at late times. Thus the only terms which survive will have additional time derivatives. It is then clear that the solution eq.(\ref{llimd}) will continue to hold  even when higher derivative corrections are included. 

However,  in the presence of higher derivative terms there could be additional solutions which do not die out at large $t$. 
We will assume that the correct boundary conditions in the far past are such that any such solution is  not ``turned on'' in the far future, leading to eq.(\ref{llimd}). 

Exponentially growing solutions would signify an instability. Our assumption  that they are absent is consistent with the background inflationary solution being stable. 
There could be additional oscillatory solutions though, which are non-decaying. We cannot rule these out except by appealing to the initial conditions. However, the following possibility is worth mentioning in this context. The additional oscillatory solutions might be present  if the higher derivative corrections in eq.(\ref{hdact}) arise  in the first place by integrating out massive particles with a mass $ \sim O(\Lambda)$. This could happen in an underlying theory where all particles, the massive ones and the graviton, satisfy second order equations of motion, leading to a well posed initial value problem. 
In this case, the graviton will indeed have the solution discussed above, eq.\eqref{llimd}, with a second solution which decays, eq.\eqref{gsold}.  If $\Lambda \sim H$, these additional particles would also be produced during inflation, with a suitable Boltzmann suppression. However, the Ward identities we derive in section \ref{identities} will continue to hold in this case as well. The wave function in the presence of these fields will continue to be invariant under spatial reparametrizations, and thus after integrating these heavy fields out, eq.\eqref{zneval1}, the same Ward identities will follow for $\zeta$ and ${\widehat \gamma}_{ij}$.

The discussion for spin-$1$ is  even more straightforward. The solution found in the two-derivative case is pure gauge, since there are no physical degrees of freedom with spin $1$. 
Starting from the unperturbed solution of the form eq.\eqref{frwback}, and carrying out a spatial reparametrization 
\be
\label{coordx}
x^i \rightarrow x^i + \epsilon^i(\vec[x]) ,
\ee
one gets 
\be
\label{valc}
C_i = \epsilon_i -\partial_i \del^{-2} (\partial\cdot \epsilon),
\ee
so that the most general time independent $C_i$ can be turned on with a suitable choice of $\epsilon^i(\vec[x])$. 
This makes it clear that a solution of the form eq.\eqref{gsolc} must continue to exist in the presence of higher derivative terms too.

Finally we come to the scalar perturbations. In the two-derivative theory, the late time behaviour for solutions was found to be eq.\eqref{gsola}, eq.\eqref{blsol} and eq.\eqref{gsoldp}, for $A, B$ and $\delta\phi$ respectively. 
We now argue, in analogy with the case of spin-$1$ above, that the existence of solutions exhibiting this behaviour follows from spatial and time reparametrizations which preserve the synchronous gauge eq.\eqref{gfix}. 
Starting from eq.\eqref{frwback} and doing the  transformation eq.\eqref{coordx} gives
\be
\label{valB}
B=2 \, \del^{-2} (\partial \cdot \epsilon),
\ee
so that the most general time independent $B$ can be turned on. 
Also, starting from eq.\eqref{frwback} and carrying out a transformation 
\be
\label{spla}
x^i \rightarrow x^i (1 + \epsilon),
\ee
where $\epsilon$ is a constant, gives 
\be
\label{valA}
A = 2 \, \epsilon .
\ee
The late time behaviour of $A$ with higher derivative terms will still be determined by an equation where all spatial derivatives can be dropped. 
The solution eq.\eqref{valA} then means that actually 
\be
\label{val2A}
A \rightarrow P_1(\vec[x])\ee
will be a solution to the small perturbation equations.

Finally, doing the time reparametrization eq.\eqref{timere} gives rise to the solution 
\be
\label{sol3}
A = 2 H \epsilon(\vec[x]), \, \delta \phi = \dot{\bar{\phi}} \, \epsilon(\vec[x]) .
\ee
Putting all these solutions together, we get the general late time behaviour seen in eq.\eqref{gsola}, eq.\eqref{blsol} and eq.\eqref{gsoldp}. 

Since these solutions in the spin-$0$ case  arise  just from gauge invariance, they will continue to be true even in the presence of the higher derivative terms. As in the case of the spin-$2$ mode, there could as well be additional solutions which do not decay, but we will assume that they are either not turned on due to the initial conditions, or are of oscillatory type  arising due to additional massive particles, which do not invalidate the arguments for the Ward identities.

\section{Conclusions}
\label{conclusions}
In this paper, we have derived the Ward identities that arise from scale and special conformal transformations for single field inflation. Our results are given in eq.\eqref{wardsc2} and eq.\eqref{wardt2} for the scalar perturbations, and eq.\eqref{wardsc4} and eq.\eqref{wardt3} for the tensor perturbations. 
Similar results for mixed correlators can also be easily obtained, see eq.\eqref{wardsc5} and eq.\eqref{finwardn}. 

The Ward identities for the special conformal transformations also involve a contribution due to a compensating spatial reparametrization, as explained in section \ref{sctr}. 
The underlying reason for this is that we are working with local correlators in a quantum theory of gravity. Such correlators can be defined in perturbation theory after suitable gauge fixing, but a compensating spatial reparametrization must then be carried out to preserve the gauge, for deriving the Ward identities of special conformal transformations \cite{Ghosh:2014kba}. 
The Ward identities for scale invariance do not require such a compensating transformation and are well known in the literature already, \cite{Maldacena:2002vr}, and called the Maldacena consistency conditions. 

The Ward identities we obtain also incorporate the breaking of the $SO(1,4)$ symmetry. In fact, this breaking is incorporated to all orders in the slow roll parameters. 
The resulting relations can be thought of as being the analogues of the Callan-Symanzik equation, but now  for both scale and special conformal transformations. 

When the slow roll conditions are approximately valid, the Ward  identities impose useful constraints on the correlation functions. The coefficient functions which appear in the wave function, and which transform in a manner analogous to correlation functions in a conformal field theory, can then be constrained  order by order in the slow approximation, and the resulting constraints on the expectation values, in agreement with the Ward identities, can then be obtained. 
For the scalar three point function, which is observationally the most important one for non-Gaussianity, this was discussed in \cite{Kundu:2014gxa}. 

We work in a theory where the degrees of freedom are  the metric and a scalar field.  However, it is worth emphasizing that our results are also valid in situations where there are extra massive fields present during inflation, with masses of order the Hubble scale, or even higher. The wave function in the presence of such fields must still meet the equations of motion imposed by varying the shift and lapse functions, and thus  must be invariant under the spatial reparametrizations discussed in section \ref{spreps}, in the gauge eq.\eqref{delphigauge} at late times. As a result, after the heavy fields are  integrated out in deriving the expectation values for $\zeta$ and ${\widehat{\gamma}_{ij}}$, in the step analogous to eq.\eqref{zneval1}, the same Ward identities as before, eq.\eqref{wardsc2}, eq.\eqref{wardsc4}, eq.\eqref{wardt2} and eq.\eqref{wardt3} are obtained.

In contrast,  our results are not valid when there are additional scalar fields which are much lighter than the Hubble scale, as in multi-field models of inflation. In this case, it is well known that the results are model dependent, and do not follow just from the underlying symmetries.
 
It will be worth examining the Ward identities of scale and special conformal transformations also in situations where the conformal symmetries are badly broken. This happens, for instance, in DBI inflation \cite{Silverstein:2003hf}, \cite{Alishahiha:2004eh}. It can also happen if the initial state is not the Bunch-Davies vacuum, and breaks the conformal symmetries to a significant extent.

In the context of AdS physics, space-times where conformal symmetry is broken are important in the study of condensed matter physics and QCD. Examples include Lifshitz and hyperscaling violating geometries. 
The Ward identities for the stress tensor etc. can be obtained in such situations in a way completely analogous to what we have used above. Some discussion of the identities in such situations can be found in \cite{Chemissany:2014xsa}.

{\textit{\underline{The Scalar Three-Point Function}}:}
Since the scalar three point function is of the greatest interest as a test of non-Gaussianity, let us end by commenting on it in some more detail. 
The Ward identities of interest here are eq.\eqref{wardsc2} and eq.\eqref{wardt2} for $n=3$, and relate the scalar $3$ and $4$ point correlators. These Ward identities were studied to the leading order in the slow roll expansion in \cite{Kundu:2014gxa}. The resulting relations in terms of coefficient functions are 
given in eq.(3.24) and eq.(3.25) of \cite{Kundu:2014gxa}, 
\begin{equation}
\label{wdscal}
\bigg( \sum_{a=1}^3 \vec[k_a] \cdot \frac{\partial}{\partial{\vec[k_a]}} \bigg) \langle O(\vecs[k,1])
O(\vecs[k,2]) O(\vecs[k,3])\rangle  = \, \frac{\dot{\bar{\phi}}}{H} \, 
\langle O(\vecs[k,1]) O(\vecs[k,2]) O(\vecs[k,3]) O(\vecs[k,4]) \rangle
\bigg\rvert_{\vecs[k,4] \rightarrow 0},
\end{equation}
and
\begin{equation}
\label{wdsct}
\begin{split}
{\cal L}^{\vec[b]}_{\vecs[k,1]} \langle O(\vecs[k,1]) O(\vecs[k,2]) O(\vecs[k,3]) \rangle' & +
{\cal L}^{\vec[b]}_{\vecs[k,2]} \langle O(\vecs[k,1]) O(\vecs[k,2]) O(\vecs[k,3]) \rangle'+
{\cal L}^{\vec[b]}_{\vecs[k,3]} \langle O(\vecs[k,1]) O(\vecs[k,2]) O(\vecs[k,3]) \rangle' \\ 
& = 2 \, \frac{\dot{\bar{\phi}}}{H} \bigg[ \vec[b] \cdot \frac{\del}{\del \vecs[k,4]} \bigg] 
\bigg\{ \langle O(\vecs[k,1]) O(\vecs[k,2]) O(\vecs[k,3]) O(\vecs[k,4]) \rangle' \bigg\rvert_{\vecs[k,4] 
\rightarrow 0} \bigg\},
\end{split}
\end{equation}
with $\mathcal{L}^{\vec[b]}_{\vec[k]}$ defined in eq.\eqref{opl}. Note that in the leading slow roll approximation, the four point coefficient function $\langle OOOO \rangle$ can be calculated in the conformally invariant limit \cite{Ghosh:2014kba}. 
As a result of the factor of ${\dot{\bar{\phi}}}$ on the RHS of eq.\eqref{wdscal} and eq.\eqref{wdsct}, the three point function $\langle OOO \rangle$ will be suppressed. Converting to expectation values, one gets that 
\be
\label{zzz}
\frac{\langle \zeta \zeta \zeta \rangle}{\langle \zeta \zeta \rangle^2} \sim \epsilon,
\ee
where the slow roll parameter $\epsilon$ is given in eq.\eqref{eps1}. 
Although the functional form one will get in general is different, this roughly corresponds to 
\be
\label{deffnl}
f_{NL}\sim \epsilon.
\ee

 It is well known that the parameter $r$ which measures the ratio of the power in the tensor to scalar perturbations is given by \footnote{The tensor and scalar power spectra $P_t(k), P_{\zeta}(k)$ are
 \begin{equation*}
  P_t(k) = \frac{H^2}{M_{Pl}^2} \frac{4}{k^3}, \; \text{and}  \; P_{\zeta}(k) = \frac{H^2}{M_{Pl}^2} \frac{1}{\epsilon} \frac{1}{4k^3}.
 \end{equation*}}
 \be
 \label{defr}
 r \equiv \frac{P_t(k)}{P_{\zeta}(k)} = 16 \epsilon.
 \ee
 
Note that in theories which are not of the type described by a canonical model of inflation, eq.\eqref{canact}, e.g., those  involving higher derivative corrections, $\epsilon_1$ and $\epsilon$ as defined in eq.\eqref{epsdef} and eq.\eqref{eps1} need not be the same. In these theories also, eq.\eqref{zzz} with the definition of $\epsilon$ given in eq.\eqref{eps1} is still valid to  leading order in the slow roll parameters. 
We see from eq.\eqref{deffnl} and eq.\eqref{defr} that there is therefore an interesting tie-in between the ratio of power in the scalar and tensor perturbations, and the non-Gaussianity. This connection is well known in the canonical slow roll models, but we see here that it is more general, since eq.\eqref{deffnl} follows from symmetry considerations alone. 

The estimate in eq.\eqref{zzz} should actually be thought of as a lower bound. A contribution due to an intermediate graviton (or the stress energy tensor running as an intermediate in the $\langle OOOO \rangle$ correlator) will give a contribution of this order to the non-Gaussianity. However, as has been emphasized in \cite{Arkani-Hamed:2015bza}, if there are additional particles of mass of order the Hubble scale which couple more strongly than the graviton, the contribution can be even bigger \footnote{Similarly, in theories where  conformal invariance is violated to a significant extent, the non-Gaussianity can be bigger, e.g. in DBI inflation.}.
 
Keeping the above  considerations in mind we can  phrase this tie-in between the two scales as follows.  If tensor perturbations are observed in the future, so that $\epsilon$ is known, we would have a firm prediction on a lower bound on non-Gaussianity that follows 
only  from conformal invariance. On the other hand, if the non-Gaussianity is observed and found to be of a bigger magnitude than  the bound on $\epsilon$  that arises from constraints on the tensor perturbations, eq.(\ref{defr}), then it would rule out the  scenario of approximate conformal invariance. More correctly, it will rule out this scenario together with the assumption that  particles which appear as intermediate states, and contribute to the non-Gaussianity, couple to the inflaton only with gravitational strength.

\section{Acknowledgements}
We thank Nima Arkani-Hamed, Juan Maldacena, Shiraz Minwalla, and Fernando Quevedo for discussions. We thank the organizers of Strings 2015 for a stimulating meeting. 
SPT thanks the DAE, Government of India, and the J. C. Bose Fellowship, DST, Govt. of India, for support.
NK thanks the organizers of the Spring School on Superstring Theory at the Abdus Salam ICTP. NK and AS thank the organizers and participants of the Advanced Strings School 2015 at ICTS Bangalore for helpful discussions.
AS would like to thank the organizers of the 32nd Winter School in Theoretical Physics at the Israel Institute of Advanced Studies, and the First ICTP Advanced School on Cosmology at the Abdus Salam ICTP, for their hospitality while this work was being done.


\appendix

\section{Transformation of Perturbations under Spatial Reparametrizations}
\label{perttrans}
In this appendix, we would like to give some details about the transformation properties of the perturbations under spatial reparametrizations. We consider the perturbed line element in the gauge eq.\eqref{gfix},
\be
\label{pertlem}
ds^2 = - \, dt^2 + h_{ij}(t,\vec[x])\, dx^i dx^j ,
\ee
with
\be
\label{genhij}
h_{ij} \equiv a^2(t) \, g_{ij} = a^2(t) \, e^{2 \zeta} \, [\delta_{ij} + \widehat\gamma_{ij}],
\ee
where
\be
\label{tracond}
\widehat\gamma_{ii} = 0.
\ee
 Consider now a spatial reparametrization of the form eq.\eqref{sprep}. The change in $h_{ij}$ under this transformation is
\be
\label{chghij}
\delta h_{ij} = \nabla_{i} v_{j} + \nabla_{j} v_{i}\,.
\ee
Eq.\eqref{chghij} implies
\begin{equation}
\begin{split}
\label{dgij}
\delta g_{ij} &= \frac{1}{a^2(t)} \left[ \del_i v_j + \del_j v_i - 2 \, \Gamma^{a}_{ij} \, v_a \right] \\
&= \frac{1}{a^2(t)} \left[ \del_i \left( h_{jk} v^k \right) + \del_j \left( h_{ik} v^k \right) - v_a \, h^{ab} \big( \del_i h_{jb} + \del_j h_{ib} - \del_b h_{ij} \big) \right] \\
&= g_{jk} \, \del_i v^k + g_{ik} \, \del_j v^k + v^k \del_k g_{ij}\, ,
\end{split}
\end{equation}
where indices will now be raised and lowered by $\delta_{ij}$. Eq.\eqref{dgij} gives us
\be
\label{dgii}
\delta g_{ii} = 2  g_{ik} \, \del_i v^k + v^k \del_k g_{ii}.
\ee
Putting $g_{ij}$ from eq.\eqref{genhij} in eq.\eqref{dgii} gives the change in $\zeta$ under spatial reparametrizations, eq.\eqref{sprep}, as
\be
\label{zchgap}
\delta \zeta = \frac{1}{3} \, \del_i v_i + v^k \del_k \zeta + \frac{1}{3} \, \del_i v_j \, \widehat\gamma_{ij} \, ,
\ee
which is the result quoted in eq.\eqref{zchgn}. Once we have calculated $\delta\zeta$, we can insert the full $g_{ij}$ in eq.\eqref{dgij} to get the change in $\widehat\gamma_{ij}$ as
\begin{equation}
\begin{split}
\label{cdgijc}
\delta \widehat\gamma_{ij} = \bigg( \del_i v_j + \del_j &v_i - \frac{2}{3} \, \del_a v_a \, \delta_{ij} \bigg) + \bigg(\widehat\gamma_{ik} \, \del_j v^k + \widehat\gamma_{jk} \, \del_i v^k + \\ &+ v^k \del_k \widehat\gamma_{ij} - \frac{2}{3} \, \del_a v_a \, \widehat\gamma_{ij} - \frac{2}{3} \, \del_a v_b \, \widehat\gamma_{ab} \left(\delta_{ij} + \widehat\gamma_{ij}\right)\bigg).
\end{split}
\end{equation}
For simplicity, we call the terms in eqs.\eqref{zchgap} and \eqref{cdgijc} which are proportional to the perturbations as the homogeneous pieces of the transformation, and the parts independent of the perturbations as the inhomogeneous pieces of the transformation. 

Having obtained eq.\eqref{zchgap} and eq.\eqref{cdgijc}, we can calculate the changes $\delta\zeta$ and $\delta\widehat{\gamma}_{ij}$ for the specific cases of scale transformations, eq.\eqref{scalechn}, special conformal transformations, eq.\eqref{sctn}, and the compensating spatial reparametrization, eq.\eqref{compens}. For scale transformations, the change in $\zeta$ and $\widehat\gamma_{ij}$ is given by eq.\eqref{zchscal1} and eq.\eqref{gchscal1} respectively. Similarly, for the special conformal transformations, the changes are given by eq.\eqref{sctzc} and eq.\eqref{sctgc}, and for the compensating spatial reparametrization, the changes are eq.\eqref{zchgcom} and eq.\eqref{cochgf}.

\section{The Scalar and Tensor Spectral Tilts}
\label{sttilts}
As a simple check on the Ward identities, we consider here the 2-point correlators. 
For scalar perturbations, the scaling Ward identity in eq.\eqref{wardsc2} relates the 2-point expectation value  to the 3-point expectation value,
\begin{equation}
\begin{split}
\label{genscm5}
\bigg( 3 + \sum_{a=1}^2 k_a \frac{\del}{\del k_a} \bigg) \langle \zeta(\vecs[k,1]) \zeta(\vecs[k,2]) \rangle' = - \, \frac{1}{\langle \zeta(\vecs[k,3]) \zeta(-\vecs[k,3])\rangle'} \, \langle \zeta(\vecs[k,1]) \zeta(\vecs[k,2]) \zeta(\vecs[k,3]) \rangle'\bigg|_{\vecs[k,3] \rightarrow 0}.
\end{split}
\end{equation}
The expression for $\langle \zeta(\vecs[k,1]) \zeta(\vecs[k,2]) \rangle$ is
\be
\label{zzdef}
\langle \zeta(\vecs[k,1]) \zeta(\vecs[k,2]) \rangle = (2\pi)^3 \delta^3(\vecs[k,1]+\vecs[k,2])\, \frac{H^2}{M_{Pl}^2} \, \frac{1}{4 \epsilon} \, k_1^{\,-3 + n_S} ,
\ee
where $n_S$ is the scalar tilt. Thus, we get
\be
\label{zzsdim}
\left[  \sum_{a = 1}^2 k_a \, \frac{\del}{\del k_a} \right] \langle \zeta(\vecs[k,1]) \zeta(\vecs[k,2]) \rangle' = (-3 + n_S) \langle \zeta(\vecs[k,1]) \zeta(-\vecs[k,1]) \rangle',
\ee
which on substituting back into the eq.\eqref{genscm5} gives the well known Maldacena consistency condition
\be
\label{mcc}
\lim_{\vecs[k,3] \rightarrow 0} \langle \zeta(\vecs[k,1]) \zeta(\vecs[k,2]) \zeta(\vecs[k,3]) \rangle' = -\, n_S \, \langle \zeta(\vecs[k,1]) \zeta(- \vecs[k,1]) \rangle' \, \langle \zeta(\vecs[k,3]) \zeta(- \vecs[k,3]) \rangle' .
\ee
By using the expression for $\langle \zeta(\vecs[k,1]) \zeta(\vecs[k,2]) \zeta(\vecs[k,3]) \rangle$ from \cite{Maldacena:2002vr}, we get\footnote{The slow roll parameter $\eta$ is given by $\eta \equiv \epsilon - \delta = \frac{V''}{V}$.}
\be
\label{limzzz}
\lim_{\vecs[k,3] \rightarrow 0} \langle \zeta(\vecs[k,1]) \zeta(\vecs[k,2]) \zeta(\vecs[k,3]) \rangle' = (6 \epsilon - 2 \eta) \, \langle \zeta(\vecs[k,1]) \zeta(- \vecs[k,1]) \rangle' \, \langle \zeta(\vecs[k,3]) \zeta(- \vecs[k,3]) \rangle' .
\ee
Putting the limit from eq.\eqref{limzzz} into eq.\eqref{mcc} gives the expression for the scalar tilt as
\be
\label{sctilt}
n_S = 2 \eta - 6 \epsilon \,,
\ee
which is indeed the correct expression, \cite{Maldacena:2002vr}.

Similarly, consider the tensor Ward identity in eq.\eqref{wardsc4}, with $n = 2$. This has the form
\begin{equation}
\label{tpthrt}
\bigg( 3 + \sum_{a=1}^2 k_a \, \frac{\del}{\del k_a} \bigg) \langle \widehat\gamma_{s}(\vecs[k,1])   \widehat\gamma_{s'}(\vecs[k,2]) \rangle' = - \, \frac{1}{\langle \zeta(\vecs[k,3]) \zeta(-\vecs[k,3])\rangle'} \, \langle \widehat\gamma_{s}(\vecs[k,1]) \widehat\gamma_{s'}(\vecs[k,2]) \zeta(\vecs[k,3])\rangle'\bigg|_{\vecs[k,3] \rightarrow 0}.
\end{equation}
In writing eq.\eqref{tpthrt}, we have introduced the two polarization tensors for the graviton, $e_{ij}^s$, through the relation
\be
\label{poltens}
\widehat\gamma_{ij}(\vec[k]) = \sum_{s = 1}^{2}  e_{ij}^s(\vec[k]) \, \widehat\gamma_{s}(\vec[k]).
\ee
Now, the expression for $\langle \widehat{\gamma}_{s}(\vecs[k,1]) \widehat{\gamma}_{s'}(\vecs[k,2]) \rangle$ has the form
\be
\label{exprgg}
\langle \widehat{\gamma}_{s}(\vecs[k,1]) \widehat{\gamma}_{s'}(\vecs[k,2]) \rangle = (2\pi)^3 \delta^3(\vecs[k,1] + \vecs[k,2]) \, \delta_{s,s'} \, \frac{H^2}{M_{Pl}^2} \, k_1^{\,-3 + n_T} ,
\ee
where $n_T$ is the tensor tilt. By using the expression eq.\eqref{exprgg} in eq.\eqref{tpthrt}, we get
\be
\label{tenscc}
\lim_{\vecs[k,3] \rightarrow 0} \, \langle \widehat{\gamma}_{s}(\vecs[k,1]) \widehat{\gamma}_{s'}(\vecs[k,2]) \zeta(\vecs[k,3]) \rangle' = -\, n_T \, \delta_{s,s'} \, \langle \widehat{\gamma}_{s}(\vecs[k,1]) \widehat{\gamma}_{s}(- \vecs[k,1]) \rangle' \langle \zeta(\vecs[k,3]) \zeta(- \vecs[k,3]) \rangle'.
\ee
We can calculate the limit on the left side of eq.\eqref{tenscc} by using the expression for the correlator $\langle \widehat{\gamma}_{s}(\vecs[k,1]) \widehat{\gamma}_{s'}(\vecs[k,2]) \zeta(\vecs[k,3]) \rangle$ from \cite{Maldacena:2002vr}. This gives
\be
\label{callim}
\lim_{\vecs[k,3] \rightarrow 0} \, \langle \widehat{\gamma}_{s}(\vecs[k,1]) \widehat{\gamma}_{s'}(\vecs[k,2]) \zeta(\vecs[k,3]) \rangle' = 2 \epsilon \, \delta_{s,s'} \langle \widehat{\gamma}_{s}(\vecs[k,1]) \widehat{\gamma}_{s}(- \vecs[k,1]) \rangle' \langle \zeta(\vecs[k,3]) \zeta(- \vecs[k,3]) \rangle'.
\ee
Then by comparing eq.\eqref{tenscc} and eq.\eqref{callim}, we get
\be
\label{tenstilt}
n_T = - \, 2 \epsilon \, ,
\ee
which is the correct expression for the tensor tilt.

\section{The Behaviour of Perturbations in Canonical Slow Roll}
\label{deteins}
In this appendix, we provide some details of the analysis given in section \ref{canmodel}. We follow \cite{Weinberg:2008zzc} for our calculations. Our gauge choice, eq.\eqref{gfix}, is same as the synchronous gauge of \cite{Weinberg:2008zzc} (see section \textbf{5.3 (B)}). The relevant equations are eqs.(5.3.28)-(5.3.33) for scalar perturbations, eq.(5.1.51) for vector perturbations, and eq.(5.1.53) for tensor perturbations.

The energy-momentum tensor for the inflaton can be calculated by varying the matter part of the action \eqref{canact} with respect to the metric. It is given by
\be
\label{defenmom}
T^{\mu \nu} = - g^{\mu \nu} \left( \half (\nabla\phi)^2 + V(\phi) \right) + g^{\mu \alpha} g^{\nu \beta} \del_{\alpha} \phi \, \del_{\beta} \phi.
\ee
This has the form of the energy-momentum tensor for a perfect fluid,
\be
\label{perffluid}
T^{\mu \nu} = (\rho + P) \, u^{\mu} u^{\nu} + P g^{\mu \nu},
\ee
with the energy density $\rho$, pressure $P$, and the four-velocity $u^{\mu}$ given by
\be
\rho = - \half (\nabla\phi)^2 + V(\phi), \label{energy}
\ee
\be
P = - \half (\nabla\phi)^2 - V(\phi), \label{pressure}
\ee
\be
u^{\mu} = - [-(\nabla\phi)^2]^{-1/2} \, g^{\mu \nu} \del_{\nu} \phi. \label{fourvelo}
\ee

For our purpose, we specialize to the case of single field slow roll inflation. We then have
\be
\bar{\rho} = \half \, \dot{\bar{\phi}}^2 + V(\bar{\phi}), \label{rhob} 
\ee
\be
\bar{P} = \half \, \dot{\bar{\phi}}^2 - V(\bar{\phi}), \label{presb}
\ee
\be
\bar{u}^0 = 1, \, \bar{u}^i = 0, \label{velob}
\ee
for the homogeneous background $\bar\phi(t)$. By expanding eqs.\eqref{energy}-\eqref{fourvelo} to linear order in the perturbation $\delta \phi$, we get
\be
\delta\rho = \dot{\bar{\phi}} \, \delta\dot{\phi} + V'(\bar{\phi}) \, \delta\phi, \label{delrho}
\ee
\be
\delta P = \dot{\bar{\phi}} \, \delta\dot{\phi} - V'(\bar{\phi}) \, \delta\phi, \label{delpres}
\ee
\be
\delta u = -\, \frac{\delta\phi}{\dot{\bar{\phi}}}, \label{delvelo}
\ee
where $\delta u$ is defined through $\delta u_i \equiv \del_i \delta u + \delta u_i^V$, and $\delta u_i^V = 0$ for single field inflation. Also, for single field inflation, the anisotropic stresses in the perturbed energy-momentum tensor vanish,
\be
\label{aniso}
\pi^S = 0, \, \pi_i^V = 0, \, \pi_{ij}^T = 0.
\ee

By using the eqs.\eqref{rhob}-\eqref{aniso} above in the eqs.(5.3.28)-(5.3.33), eq.(5.1.51) and eq.(5.1.53) of \cite{Weinberg:2008zzc}, we obtain the perturbed Einstein equations eqs.\eqref{eqa}-\eqref{eomdp} given in section \ref{canmodel} for the scalar, vector and tensor perturbations, along with the equation of motion for the background $\bar\phi(t)$, eq.\eqref{bck}. Note that the perturbations $G_j$ in eq.(5.1.51) of \cite{Weinberg:2008zzc} vanish due to our gauge choice eq.\eqref{gfix}.

We now provide some details for calculating the late time behaviour of the perturbations. To solve for the perturbation $A$, we consider eq.\eqref{eqg}. Inserting $\delta\phi$ from eq.\eqref{eqc} into eq.\eqref{eqg}, we get an equation purely for the perturbation $A$,
\be
\label{eqnaal}
\frac{\ddot{A}}{2} + \left[ 3 \left( \frac{\dot{a}}{a}\right) + \frac{V'(\bar{\phi})}{\dot{\bar{\phi}}} \right] \dot{A} = 0.
\ee
By using the background eq.\eqref{bck} in eq.\eqref{eqnaal}, we get
\be
\label{eqnal}
\ddot{A} - 2 \left( \frac{\ddot{\bar{\phi}}}{\dot{\bar{\phi}}} \right) \dot{A} = 0.
\ee
The general solution to eq.\eqref{eqnal} is
\be
\label{gsolap}
A(t,\vec[x]) = P_1(\vec[x])+ P_2(\vec[x]) \int^{\,t} dt' \, \dot{\bar{\phi}}^{\,2}(t') ,
\ee
where $P_1(\vec[x]), P_2(\vec[x])$ are two arbitrary functions of $\vec[x]$. Eq.\eqref{gsolap} on using the background equation
\be
\label{bckhp}
\dot{H} \equiv \frac{d}{dt}\left( \frac{\dot{a}}{a}\right) = - \, \half \, \dot{\bar{\phi}}^{\,2}
\ee
becomes
\begin{equation*}
A(t,\vec[x]) = P_1(\vec[x]) - 2 \left( \frac{\dot{a}}{a}\right) P_2(\vec[x]),
\end{equation*}
which is the solution quoted in eq.\eqref{gsola}.

Once we have obtained the solution for $A$, it is straight forward to obtain the solution for the perturbation $\delta\phi$. From eq.\eqref{eqc} and eq.\eqref{gsola}, it follows that
\begin{equation*}
\delta\phi(t,\vec[x]) = -\, \dot{\bar{\phi}}(t) \, P_2(\vec[x]),
\end{equation*}
as given in eq.\eqref{gsoldp}. One can check explicitly that the solutions eq.\eqref{gsola}, eq.\eqref{gsoldp} satisfy the other equations, namely eq.\eqref{eqh} and eq.\eqref{eqj}.

The equation for the perturbation $C_i$, eq.\eqref{eqe}, has the general solution
\be
\label{gsolc}
C_i(t,\vec[x]) =  \del^{-2} Q_i(\vec[x]),
\ee
which shows that the perturbation $C_i$ is frozen for non-zero momentum modes, which are the ones of interest to us.

Finally, we consider eq.\eqref{eqi} for the tensor perturbations. The general solution is
\be
\label{gsold}
D_{ij}(t,\vec[x]) = \tilde{D}_{ij}(\vec[x]) + K_{ij}(\vec[x]) \int^{\,t} dt' \, \text{exp}\left[ - \,3 \int^{\,t'} dt'' \, \left( \frac{\dot{a}}{a}\right) \right],
\ee
which in the late time limit also gets frozen,
\be
\label{llimd}
D_{ij}(t,\vec[x]) \approx \tilde{D}_{ij}(\vec[x]) \; \text{for t} \rightarrow \infty.
\ee


\bibliography{references}{}

\providecommand{\href}[2]{#2}\begingroup\raggedright\begin{thebibliography}{10}

\bibitem{Maldacena:2002vr}
J.~M. Maldacena, {\it {Non-Gaussian features of primordial fluctuations in
  single field inflationary models}},  {\em JHEP} {\bf 0305} (2003) 013,
  [\href{http://arxiv.org/abs/astro-ph/0210603}{{\tt astro-ph/0210603}}].

\bibitem{Maldacena:2011nz}
J.~M. Maldacena and G.~L. Pimentel, {\it {On graviton non-Gaussianities during
  inflation}},  {\em JHEP} {\bf 1109} (2011) 045,
  [\href{http://arxiv.org/abs/1104.2846}{{\tt arXiv:1104.2846}}].

\bibitem{Mata:2012bx}
I.~Mata, S.~Raju, and S.~P. Trivedi, {\it {CMB from CFT}},  {\em JHEP} {\bf
  1307} (2013) 015, [\href{http://arxiv.org/abs/1211.5482}{{\tt
  arXiv:1211.5482}}].

\bibitem{Ghosh:2014kba}
A.~Ghosh, N.~Kundu, S.~Raju, and S.~P. Trivedi, {\it {Conformal Invariance and
  the Four Point Scalar Correlator in Slow-Roll Inflation}},  {\em JHEP} {\bf
  1407} (2014) 011, [\href{http://arxiv.org/abs/1401.1426}{{\tt
  arXiv:1401.1426}}].

\bibitem{Kundu:2014gxa}
N.~Kundu, A.~Shukla, and S.~P. Trivedi, {\it {Constraints from Conformal
  Symmetry on the Three Point Scalar Correlator in Inflation}},  {\em JHEP}
  {\bf 1504} (2015) 061, [\href{http://arxiv.org/abs/1410.2606}{{\tt
  arXiv:1410.2606}}].

\bibitem{Antoniadis:1996dj}
I.~Antoniadis, P.~O. Mazur, and E.~Mottola, {\it {Conformal invariance and
  cosmic background radiation}},  {\em Phys.Rev.Lett.} {\bf 79} (1997) 14--17,
  [\href{http://arxiv.org/abs/astro-ph/9611208}{{\tt astro-ph/9611208}}].

\bibitem{Larsen:2002et}
F.~Larsen, J.~P. van~der Schaar, and R.~G. Leigh, {\it {De Sitter holography
  and the cosmic microwave background}},  {\em JHEP} {\bf 0204} (2002) 047,
  [\href{http://arxiv.org/abs/hep-th/0202127}{{\tt hep-th/0202127}}].

\bibitem{Larsen:2003pf}
F.~Larsen and R.~McNees, {\it {Inflation and de Sitter holography}},  {\em
  JHEP} {\bf 0307} (2003) 051, [\href{http://arxiv.org/abs/hep-th/0307026}{{\tt
  hep-th/0307026}}].

\bibitem{McFadden:2010vh}
P.~McFadden and K.~Skenderis, {\it {Holographic Non-Gaussianity}},  {\em JCAP}
  {\bf 1105} (2011) 013, [\href{http://arxiv.org/abs/1011.0452}{{\tt
  arXiv:1011.0452}}].

\bibitem{Antoniadis:2011ib}
I.~Antoniadis, P.~O. Mazur, and E.~Mottola, {\it {Conformal Invariance, Dark
  Energy, and CMB Non-Gaussianity}},  {\em JCAP} {\bf 1209} (2012) 024,
  [\href{http://arxiv.org/abs/1103.4164}{{\tt arXiv:1103.4164}}].

\bibitem{McFadden:2011kk}
P.~McFadden and K.~Skenderis, {\it {Cosmological 3-point correlators from
  holography}},  {\em JCAP} {\bf 1106} (2011) 030,
  [\href{http://arxiv.org/abs/1104.3894}{{\tt arXiv:1104.3894}}].

\bibitem{Creminelli:2011mw}
P.~Creminelli, {\it {Conformal invariance of scalar perturbations in
  inflation}},  {\em Phys.Rev.} {\bf D85} (2012) 041302,
  [\href{http://arxiv.org/abs/1108.0874}{{\tt arXiv:1108.0874}}].

\bibitem{Bzowski:2011ab}
A.~Bzowski, P.~McFadden, and K.~Skenderis, {\it {Holographic predictions for
  cosmological 3-point functions}},  {\em JHEP} {\bf 1203} (2012) 091,
  [\href{http://arxiv.org/abs/1112.1967}{{\tt arXiv:1112.1967}}].

\bibitem{Kehagias:2012pd}
A.~Kehagias and A.~Riotto, {\it {Operator Product Expansion of Inflationary
  Correlators and Conformal Symmetry of de Sitter}},  {\em Nucl.Phys.} {\bf
  B864} (2012) 492--529, [\href{http://arxiv.org/abs/1205.1523}{{\tt
  arXiv:1205.1523}}].

\bibitem{Kehagias:2012td}
A.~Kehagias and A.~Riotto, {\it {The Four-point Correlator in Multifield
  Inflation, the Operator Product Expansion and the Symmetries of de Sitter}},
  {\em Nucl.Phys.} {\bf B868} (2013) 577--595,
  [\href{http://arxiv.org/abs/1210.1918}{{\tt arXiv:1210.1918}}].

\bibitem{Schalm:2012pi}
K.~Schalm, G.~Shiu, and T.~van~der Aalst, {\it {Consistency condition for
  inflation from (broken) conformal symmetry}},  {\em JCAP} {\bf 1303} (2013)
  005, [\href{http://arxiv.org/abs/1211.2157}{{\tt arXiv:1211.2157}}].

\bibitem{Bzowski:2012ih}
A.~Bzowski, P.~McFadden, and K.~Skenderis, {\it {Holography for inflation using
  conformal perturbation theory}},  {\em JHEP} {\bf 1304} (2013) 047,
  [\href{http://arxiv.org/abs/1211.4550}{{\tt arXiv:1211.4550}}].

\bibitem{McFadden:2014nta}
P.~McFadden, {\it {Soft limits in holographic cosmology}},  {\em JHEP} {\bf 02}
  (2015) 053, [\href{http://arxiv.org/abs/1412.1874}{{\tt arXiv:1412.1874}}].

\bibitem{Kehagias:2015jha}
A.~Kehagias and A.~Riotto, {\it {High Energy Physics Signatures from Inflation
  and Conformal Symmetry of de Sitter}},  {\em Fortsch. Phys.} {\bf 63} (2015)
  531--542, [\href{http://arxiv.org/abs/1501.03515}{{\tt arXiv:1501.03515}}].

\bibitem{Weinberg:2003sw}
S.~Weinberg, {\it {Adiabatic modes in cosmology}},  {\em Phys.Rev.} {\bf D67}
  (2003) 123504, [\href{http://arxiv.org/abs/astro-ph/0302326}{{\tt
  astro-ph/0302326}}].

\bibitem{Creminelli:2004yq}
P.~Creminelli and M.~Zaldarriaga, {\it {Single field consistency relation for
  the 3-point function}},  {\em JCAP} {\bf 0410} (2004) 006,
  [\href{http://arxiv.org/abs/astro-ph/0407059}{{\tt astro-ph/0407059}}].

\bibitem{Cheung:2007sv}
C.~Cheung, A.~L. Fitzpatrick, J.~Kaplan, and L.~Senatore, {\it {On the
  consistency relation of the 3-point function in single field inflation}},
  {\em JCAP} {\bf 0802} (2008) 021, [\href{http://arxiv.org/abs/0709.0295}{{\tt
  arXiv:0709.0295}}].

\bibitem{Weinberg:2008nf}
S.~Weinberg, {\it {Non-Gaussian Correlations Outside the Horizon}},  {\em
  Phys.Rev.} {\bf D78} (2008) 123521,
  [\href{http://arxiv.org/abs/0808.2909}{{\tt arXiv:0808.2909}}].

\bibitem{Creminelli:2011sq}
P.~Creminelli, C.~Pitrou, and F.~Vernizzi, {\it {The CMB bispectrum in the
  squeezed limit}},  {\em JCAP} {\bf 1111} (2011) 025,
  [\href{http://arxiv.org/abs/1109.1822}{{\tt arXiv:1109.1822}}].

\bibitem{Bartolo:2011wb}
N.~Bartolo, S.~Matarrese, and A.~Riotto, {\it {Non-Gaussianity in the Cosmic
  Microwave Background Anisotropies at Recombination in the Squeezed limit}},
  {\em JCAP} {\bf 1202} (2012) 017, [\href{http://arxiv.org/abs/1109.2043}{{\tt
  arXiv:1109.2043}}].

\bibitem{Creminelli:2012ed}
P.~Creminelli, J.~Norena, and M.~Simonovic, {\it {Conformal consistency
  relations for single-field inflation}},  {\em JCAP} {\bf 1207} (2012) 052,
  [\href{http://arxiv.org/abs/1203.4595}{{\tt arXiv:1203.4595}}].

\bibitem{Hinterbichler:2012nm}
K.~Hinterbichler, L.~Hui, and J.~Khoury, {\it {Conformal Symmetries of
  Adiabatic Modes in Cosmology}},  {\em JCAP} {\bf 1208} (2012) 017,
  [\href{http://arxiv.org/abs/1203.6351}{{\tt arXiv:1203.6351}}].

\bibitem{Senatore:2012wy}
L.~Senatore and M.~Zaldarriaga, {\it {A Note on the Consistency Condition of
  Primordial Fluctuations}},  {\em JCAP} {\bf 1208} (2012) 001,
  [\href{http://arxiv.org/abs/1203.6884}{{\tt arXiv:1203.6884}}].

\bibitem{Assassi:2012zq}
V.~Assassi, D.~Baumann, and D.~Green, {\it {On Soft Limits of Inflationary
  Correlation Functions}},  {\em JCAP} {\bf 1211} (2012) 047,
  [\href{http://arxiv.org/abs/1204.4207}{{\tt arXiv:1204.4207}}].

\bibitem{Creminelli:2012qr}
P.~Creminelli, A.~Joyce, J.~Khoury, and M.~Simonovic, {\it {Consistency
  Relations for the Conformal Mechanism}},  {\em JCAP} {\bf 1304} (2013) 020,
  [\href{http://arxiv.org/abs/1212.3329}{{\tt arXiv:1212.3329}}].

\bibitem{Goldberger:2013rsa}
W.~D. Goldberger, L.~Hui, and A.~Nicolis, {\it {One-particle-irreducible
  consistency relations for cosmological perturbations}},  {\em Phys.Rev.} {\bf
  D87} (2013) 103520, [\href{http://arxiv.org/abs/1303.1193}{{\tt
  arXiv:1303.1193}}].

\bibitem{Hinterbichler:2013dpa}
K.~Hinterbichler, L.~Hui, and J.~Khoury, {\it {An Infinite Set of Ward
  Identities for Adiabatic Modes in Cosmology}},
  \href{http://arxiv.org/abs/1304.5527}{{\tt arXiv:1304.5527}}.

\bibitem{Creminelli:2013cga}
P.~Creminelli, A.~Perko, L.~Senatore, M.~Simonovic, and G.~Trevisan, {\it {The
  Physical Squeezed Limit: Consistency Relations at Order $q^2$}},  {\em JCAP}
  {\bf 1311} (2013) 015, [\href{http://arxiv.org/abs/1307.0503}{{\tt
  arXiv:1307.0503}}].

\bibitem{Pimentel:2013gza}
G.~L. Pimentel, {\it {Inflationary Consistency Conditions from a Wavefunctional
  Perspective}},  {\em JHEP} {\bf 1402} (2014) 124,
  [\href{http://arxiv.org/abs/1309.1793}{{\tt arXiv:1309.1793}}].

\bibitem{Berezhiani:2013ewa}
L.~Berezhiani and J.~Khoury, {\it {Slavnov-Taylor Identities for Primordial
  Perturbations}},  \href{http://arxiv.org/abs/1309.4461}{{\tt
  arXiv:1309.4461}}.

\bibitem{Sreenath:2014nka}
V.~Sreenath and L.~Sriramkumar, {\it {Examining the consistency relations
  describing the three-point functions involving tensors}},  {\em JCAP} {\bf
  1410} (2014), no.~10 021, [\href{http://arxiv.org/abs/1406.1609}{{\tt
  arXiv:1406.1609}}].

\bibitem{Mirbabayi:2014zpa}
M.~Mirbabayi and M.~Zaldarriaga, {\it {Double Soft Limits of Cosmological
  Correlations}},  \href{http://arxiv.org/abs/1409.6317}{{\tt
  arXiv:1409.6317}}.

\bibitem{Joyce:2014aqa}
A.~Joyce, J.~Khoury, and M.~Simonovic, {\it {Multiple Soft Limits of
  Cosmological Correlation Functions}},
  \href{http://arxiv.org/abs/1409.6318}{{\tt arXiv:1409.6318}}.

\bibitem{Sreenath:2014nca}
V.~Sreenath, D.~K. Hazra, and L.~Sriramkumar, {\it {On the scalar consistency
  relation away from slow roll}},  \href{http://arxiv.org/abs/1410.0252}{{\tt
  arXiv:1410.0252}}.

\bibitem{deBoer:1999xf}
J.~de~Boer, E.~P. Verlinde, and H.~L. Verlinde, {\it {On the holographic
  renormalization group}},  {\em JHEP} {\bf 0008} (2000) 003,
  [\href{http://arxiv.org/abs/hep-th/9912012}{{\tt hep-th/9912012}}].

\bibitem{Weinberg:2008zzc}
S.~Weinberg, {\em {Cosmology}}.
\newblock Oxford University Press, 2008.

\bibitem{Silverstein:2003hf}
E.~Silverstein and D.~Tong, {\it {Scalar speed limits and cosmology:
  Acceleration from D-cceleration}},  {\em Phys.Rev.} {\bf D70} (2004) 103505,
  [\href{http://arxiv.org/abs/hep-th/0310221}{{\tt hep-th/0310221}}].

\bibitem{Alishahiha:2004eh}
M.~Alishahiha, E.~Silverstein, and D.~Tong, {\it {DBI in the sky}},  {\em
  Phys.Rev.} {\bf D70} (2004) 123505,
  [\href{http://arxiv.org/abs/hep-th/0404084}{{\tt hep-th/0404084}}].

\bibitem{Chemissany:2014xsa}
W.~Chemissany and I.~Papadimitriou, {\it {Lifshitz holography: The whole
  shebang}},  \href{http://arxiv.org/abs/1408.0795}{{\tt arXiv:1408.0795}}.

\bibitem{Arkani-Hamed:2015bza}
N.~Arkani-Hamed and J.~Maldacena, {\it {Cosmological Collider Physics}},
  \href{http://arxiv.org/abs/1503.08043}{{\tt arXiv:1503.08043}}.

\end{thebibliography}\endgroup
\bibliographystyle{JHEP}
\end{document}